\documentstyle[preprint,aps]{revtex}
\newcommand{\be}{\begin{equation}}
\newcommand{\ee}{\end{equation}}
\newcommand{\bea}{\begin{eqnarray}}
\newcommand{\eea}{\end{eqnarray}}
\begin{document}
\draft
\title{
Temporal Forcing of Small-Amplitude Waves in Anisotropic Systems
}
\author{Hermann Riecke$^+$}
\address{
Department of Engineering Sciences and Applied Mathematics \\
Northwestern University, Evanston, IL 60208, USA
}
\author{Mary Silber$^\#$}
\address{
Department of Applied Mechanics\\
Caltech, Pasadena, CA 91125, USA
}
\author{Lorenz Kramer$^+$}
\address{
Physikalisches Institut\\
Universit\"at Bayreuth, D-8580 Bayreuth, Germany
}

\maketitle

\begin{abstract}

We investigate the effect of resonant temporal forcing on an
anisotropic system that exhibits a Hopf bifurcation to obliquely
traveling waves in the absence of this forcing. We find that the
forcing can excite various phase-locked standing-wave structures:
rolls, rectangles and cross rolls.  At onset, at most one of the two -
rolls or rectangles - is stable. The cross rolls can arise in a
secondary bifurcation and can be stable. Experimentally, they would
appear as a periodic switching between a structure in which the
`zig'-component dominates and one with dominating `zag'-structure.
Since there are two symmetry-related states of this kind one may
expect disordered structures to arise due to the break-up of the
pattern into domains. The results are consistent with recent
experiments on electro-convection in nematic liquid crystals by de la
Torre and Rehberg.  We also apply the general analysis to a model of
the behavior near a Lifshitz point, where the angle of obliqueness
vanishes. This analysis indicates that phase-locked standing
rectangles are always unstable in this parameter regime.

\end{abstract}
\bigskip

\today
\bigskip
\bigskip
\bigskip
\bigskip

\pacs{PACS numbers: 47.20.Ky, 03.40.Kf, 47.25.Qv}

\narrowtext
\section{Introduction}

Temporally-periodic forcing can qualitatively change the behavior of
systems exhibiting spatial or spatio-temporal structures. In systems
which form oscillatory structures, strongly resonant temporal forcing
has a particularly large impact. For example, in one-dimensional
systems that undergo a Hopf bifurcation to stable traveling waves,
forcing can excite stable standing waves below the threshold for the
traveling waves; these standing waves are phase-locked to the forcing
\cite{rck88,w88,rck91}. This theoretical prediction has been confirmed
experimentally in Taylor-Dean flow \cite{apriv} and in
electro-convection of nematic liquid crystals (EHC) \cite{rrfts88}.

In the nematic phase, the liquid crystal system exhibits an axial
anisotropy due to a partial alignment of the molecules; this defines
the director. In the first EHC experiments the waves always propagated
along the director \cite{rrfts88}. Recently the experimental parameter
regime has been extended to include a regime in which disordered
patches of waves, traveling at an {\it oblique} angle to the director,
are observed \cite{tr90}. In this case, temporally-periodic forcing
excites standing waves in the form of either oblique rolls or
rectangles. In addition, more complicated structures are found
\cite{tr90}. These observations provide the experimental
motivation for our theoretical investigation.

This paper examines the effect of resonant forcing on oblique
traveling waves produced in a Hopf bifurcation of the spatially
uniform state of a two-dimensional anisotropic system. The symmetry of
the system forces coexistence of waves traveling in four directions.
The possible interactions of these waves makes a complete analysis of
the problem extremely complicated. However, one of the striking
features of the excited roll and rectangular structures is their
strong spatial coherence. This suggests that as a first step we focus
on the case where the amplitudes of the different modes are
space-independent. We therefore investigate an extension of the four
complex amplitude equations analyzed previously
\cite{srk92}; this model takes the temporal forcing into account
\cite{z91}, but neglects large scale spatial variations in the amplitudes.
This analysis elucidates the connection between the different
structures excited in the experiments and allows us to make some
predictions for future experiments. We also consider the transition
from oblique to normal traveling waves using suitably extended coupled
Ginzburg-Landau equations which are valid close to a Lifshitz point
where the angle of obliqueness goes to zero.

Formally, our analysis applies to weakly forced waves arising through
a Hopf bifurcation. We expect, however, that some aspects may pertain
to parametrically excited waves in systems that do not undergo a Hopf
bifurcation, but which support weakly damped waves in the absence of
forcing. In particular, this approach has been successfully applied to
surface waves arising in the Faraday experiment with low-viscosity
liquids
\cite{errs86} and is related to methods used in studies of spin
waves in ferromagnets \cite{sz86}.
Of course, when applying the
present approach to these systems only the phenomena arising in the
linearly damped regime are of relevance ({\it i.e.}, for parameter
values below the Hopf bifurcation point).

The organization of the paper is as follows. Section II presents the
amplitude equations for weakly forced oblique waves. Section III
reviews and interprets results on the transition from traveling waves
to phase-locked standing waves. The phase-locked structures, as well
as the transitions between them, are discussed in detail in section
IV. Section V contains our analysis of the Lifshitz equations. The
theoretical results are then compared with the experimental ones in
section VI. The main results of the paper are summarized in the
conclusion, section VII.

\section{Amplitude Equations for Resonantly Forced Oblique Waves}

This paper focuses on Hopf bifurcation in axially anisotropic systems
in the case where there are four neutrally stable modes at the Hopf
bifurcation point. These modes correspond to waves that travel at an
oblique angle $\gamma$ to the axis of anisotropy. Thus, a typical
scalar field such as the vertical component of the velocity in convection,
$V_z$, is given by
\bea
V_z(x,y,t)=\left(z_1 e^{i(qx+py)}+z_2 e^{i(-qx+py)}+z_3
e^{-i(qx+py)}+z_4 e^{-i(-qx+py)}\right)e^{i\omega_h t} + c.c. +
h.o.t., \label{e:vz}
\eea
where we have suppressed any dependence of $V_z$ on the vertical
coordinate $z$, and where $\omega_h$ is the Hopf frequency. Note that
$z_1$ and $z_3$ correspond to left- and right-traveling plane waves in
the direction of $q\hat {\bf x}+p\hat {\bf y}$; similarly $z_2$ and
$z_4$ correspond to oppositely traveling plane waves in the direction
of $-q\hat {\bf x}+p\hat {\bf y}$. Thus, the waves travel at an
oblique angle $\gamma=\tan ^{-1}(p/q)$ to the $x$-axis (the axis of
anisotropy). Here we assume that the amplitudes $z_i$ are
complex-valued functions of a slow time $T$, but that they do not
depend on the spatial coordinates.  The time evolution of the
amplitudes is determined by a system of coupled first order ordinary
differential equations which can, in principle, be derived from the
governing hydrodynamic equations (with periodic boundary conditions)
using center-manifold reduction \cite{gh}.  However, in EHC this is
not yet possible; even the origin of the Hopf bifurcation has not been
understood at the hydrodynamic level
\cite{kbptz89}. Nevertheless, the general form of the amplitude equations can
be determined from symmetry considerations assuming that the waves are
due to a Hopf bifurcation of the motionless state as experiments
suggest
\cite{rrs89}. Consequently, most of our results are not specific to
the EHC-experiments, but pertain to a larger class of resonantly
forced systems with the same underlying symmetries. In a more
heuristic fashion, we also expect our results to be relevant to
surface waves in the Faraday experiment
\cite{fn2}. We will, however, focus on the EHC-system in order
to make contact with available experiments \cite{tr90}.

We consider a general system that is invariant under translations in
the plane; since our solutions are spatially doubly-periodic, the
group of translations corresponds to a torus $T^2$. Moreover, we
assume reflection symmetries in planes parallel and perpendicular to
the axis of anisotropy; these reflections are denoted by $\kappa_1$ and
$\kappa_2$, where $\kappa_1:(x,y)\to (x,-y)$ and $\kappa_2:(x,y)\to
(-x,y)$. These spatial symmetries act on the amplitudes as follows:
\begin{equation}(\theta_1,\theta_2)\cdot\left(\begin{array}{c}
z_1\\
z_2\\
z_3\\
z_4\end{array}
\right)\rightarrow\left(\begin{array}{c}
e^{_{i\theta_1}}z_1\\
e^{i\theta_2}z_2\\
e^{-i\theta_1}z_3\\
e^{-i\theta_2}z_4\end{array}
\right),\ \kappa_1\cdot\left(\begin{array}{c}
z_1\\
z_2\\
z_3\\
z_4\end{array}
\right)\rightarrow\left(\begin{array}{c}
z_4\\
z_3\\
z_2\\
z_1\end{array}
\right),\ \kappa_2\cdot\left(\begin{array}{c}
z_1\\
z_2\\
z_3\\
z_4\end{array}
\right)\rightarrow\left(\begin{array}{c}
z_2\\
z_1\\
z_4\\
z_3\end{array}
\right).\label{e:action}\end{equation}
Here the phase shift $\theta_1$ results from a translation $(x,y)\to
(x+\theta_1/2q,y+\theta_1/2p)$ and the $\theta_2$ phase shift results
from a translation by $(-\theta_2/2q,\theta_2/2p)$.  The amplitude
equations are equivariant with respect to these transformations. In
the absence of resonant temporal forcing the normal form of the
amplitude equations possesses an additional phase shift symmetry that
we interpret as a time-translation symmetry $t\to t+\phi/\omega_h$. It acts on
the amplitudes as follows:
\begin{equation}
\phi\cdot{\bf z}=e^{i\phi}{\bf z},\label{e:s1}
\end{equation}
where ${\bf z}\equiv (z_1,z_2,z_3,z_4).$
An analysis of the cubic truncation of the amplitude equations,
equivariant with respect to the symmetries (\ref{e:action}) and
(\ref{e:s1}), is presented in \cite{srk92}.

The temporal forcing destroys the phase-shift symmetry (\ref{e:s1}),
thereby allowing additional terms in the amplitude equations. The
frequency of the external forcing is
$\omega_e=n\omega_h+\Delta\omega$, where we assume that the detuning
$\Delta\omega$ is small and that $n=1$ or $2$ ($n=2$ in the EHC
experiments). Moreover, we assume that the temporal forcing is weak
and therefore we keep only the lowest order term which does not
satisfy the time-translation symmetry (\ref{e:s1}).  The amplitude
equations truncated at cubic order are \cite{z91}
\begin{eqnarray}
\dot z_1 &=& \nu z_1+\mu z_3^*+\left(a\vert z_1\vert^2
+b\vert z_2\vert^2+c \vert z_3\vert^2
+d\vert z_4\vert^2\right)z_1+f\
z_2z_3^*z_4\nonumber \\
\dot z_2 &=& \nu z_2+\mu z_4^*+\left(a\vert z_2\vert^2+
b\vert z_1\vert^2+c\vert z_4\vert^2
+d\vert z_3\vert^2\right)z_2+f\
z_1z_4^*z_3\nonumber \\
\dot z_3 &=& \nu z_3+\mu z_1^*+\left(a\vert z_3\vert^2+
b\vert z_4\vert^2+c \vert z_1\vert^2
+d \vert z_2\vert^2\right)z_3+f\
z_2z_1^*z_4\label{e:agl3} \\
\dot z_4 &=& \nu z_4+\mu z_2^*+\left(a\vert z_4\vert^2+
b\vert z_3\vert^2+c \vert z_2\vert^2
+d \vert z_1\vert^2\right)z_4+f\
z_1z_2^*z_3 \ .\nonumber
\end{eqnarray}
There are three small unfolding parameters: the distance from the
Hopf bifurcation in the absence of forcing, $Re\{ \nu \}
\equiv \nu_r$, the detuning $\Delta \omega$ of the external frequency
with respect to the Hopf frequency, $Im\{ \nu \} \equiv \nu_i \propto
\Delta \omega$, and the forcing amplitude $F$, $|\mu|^n\propto F^2$
\cite{rck88,w88}. (In the case of parametrically excited, weakly damped
waves, $\nu_r$ is proportional to the linear damping of the waves; it
is always negative.) Without loss of generality, we choose the
temporal phase of the forcing such that $\mu$ is real and positive;
all other coefficients are complex.  Finally, we replace $\omega_h$ by
$\omega_e/n$ in the expression (\ref{e:vz}) for the velocity component
$V_z$.

In the absence of forcing ($\mu=0$) eqs. (\ref{e:agl3}) possess up to
seven branches of periodic solutions that bifurcate from the origin at
$\nu_r=0$. Three of these correspond to different types of traveling
waves and four to (superpositions) of standing waves. In particular,
they have the form
\begin{eqnarray}
I:\hspace{.5cm}&{\bf z}=(v,0,0,0)\hspace{.5cm}&TRo,\nonumber\\
II:&{\bf z}=(v,v,0,0) &TRe^{\perp},\nonumber\\
III:&{\bf z}=(v,0,0,v) &TRe^{\parallel},\nonumber\\
IV:&{\bf z}=(v,0,v,0) &SRo,\label{oldfps}\\
V:&{\bf z}=(v,v,v,v) &SRe,\nonumber \\
VI:&{\bf z}=(v,iv,v,iv) &ARo,\nonumber \\
VII:&{\bf z}=(v_1,v_2,v_1,v_2) &SCR.\nonumber
\end{eqnarray}
The abbreviations denote traveling rolls ($TRo$), two kinds of
traveling rectangles distinguished by their direction of propagation
relative to the axis of anisotropy $(TRe^{\perp},TRe^{\parallel})$,
standing rolls ($SRo$), standing rectangles ($SRe$), alternating rolls
($ARo$) and standing cross rolls ($SCR$). The standing
cross rolls, which are characterized by two complex amplitudes $v_1$
and $v_2$, do not always exist. The numbering is the same as that used
in our previous work on the unforced system
\cite{srk92}. In addition to these periodic solutions one finds more
complicated (branches of) attractors such as heteroclinic cycles
connecting three of the above solutions \cite{srk92}.

In the presence of forcing, oppositely traveling waves are linearly
coupled (see eq. (\ref{e:agl3})) so that pure traveling waves do not
exist. For example, the forcing destroys pure right-traveling
$TRe^{\parallel}$ by coupling it to the left-traveling
$TRe^{\parallel}$; thus the $TRe^\parallel$ are replaced by a general
superposition of left- and right-traveling rectangles, {\it i.e.} by
$(v_1,v_2,v_2,v_1)$. Similarly, forcing destroys the $ARo$ solution,
which is an equal-amplitude superposition of standing rolls which
differ by $\pi/2$ in their temporal phase. In the next section we see
that the pure standing waves ($SRo$ and $SRe$) persist and that they
can phase-lock to the temporally-periodic forcing.

\section{Traveling {\it vs. } Phase-locked Standing Waves}
\label{s:trsw}

In this section we study the effects of resonant forcing on
competition between various traveling-wave structures and their
standing-wave counterparts. Specifically, we focus on the stability of
the phase-locked standing waves to traveling-wave disturbances. In the
oblique regime there are two simple types of phase-locked standing
waves ($PSW$): phase-locked standing rolls ($PSRo$) and phase-locked
standing rectangles ($PSRe$). The stability of the $PSRo$ state to
traveling disturbances is determined by restricting (\ref{e:agl3}) to
the subspace where ${\bf z}=(v_1,0,v_2,0)$.  Similarly the stability
of $PSRe$ to traveling disturbances is determined by restricting
(\ref{e:agl3}), in turn, to each of the subspaces where ${\bf
z}=(v_1,v_1,v_2,v_2)$ and ${\bf z}=(v_1,v_2,v_2,v_1)$; in the former
case the disturbances travel perpendicular to the director, whereas in
the latter case they travel along it. In each of these three cases,
the restricted dynamics has the form
\bea
\dot {v}_1=\nu v_1+\mu v_2^{*}+\alpha v_1|v_1|^2+\beta v_1|v_2|^2,
\nonumber\\
\dot {v}_2=\nu v_2+\mu v_1^{*}+\beta v_2|v_1|^2+\alpha v_2|v_2|^2,
\label{e:modhopf}
\eea
where
\begin{mathletters}
\bea
\alpha=a, \quad \quad \quad & \beta=c\quad \quad \quad \quad &\mbox{  for }
TRo,\label{e:1d2da}
\\
\alpha=a+b, \quad \quad \quad& \beta=c+d+f\quad &\mbox{  for } TRe^\perp
\label{e:1d2db},\\
\alpha=a+d, \quad \quad \quad
& \beta=b+c+f &\quad\mbox{  for } TRe^\parallel. \label{e:1d2dc}
\eea
\end{mathletters}
Here the subspaces are denoted by the traveling waves that are present
in each of them when there is no resonant forcing ({\it cf.}
(\ref{oldfps})). Eqn. (\ref{e:modhopf}) was studied previously in the
context of periodically forced waves in one dimension
\cite{rck88,w88,rck91}; in these papers the phase-locked
standing waves ($PSW$), as well as traveling ($TW$) and unlocked
standing waves ($SW$) were investigated. The results apply directly to the
present
system by an appropriate identification of $TRo$ and
$TRe^{\parallel,\perp}$ with the traveling wave $TW$, and the new
phase-locked solutions $PSRo$ and $PSRe$ with the standing wave $PSW$.

In fig.\ref{f:pdsup} we sketch a typical phase diagram associated with
(\ref{e:modhopf}), for fixed detuning $\nu_i$, in the case that the
unforced traveling waves bifurcate supercritically and are stable.  In
the following discussion we employ the notation ($TW$, $PSW$) used in
the one-dimensional case.  For weak forcing the trivial equilibrium
loses stability via Hopf bifurcation along the line marked $H$; this
bifurcation leads to stable traveling waves, $TW$, for positive values
of $\nu_r$. (The unlocked standing waves, $SW$, are unstable in this
case.) Along the line marked $PSW$, which is given by $\mu^2=|\nu|^2$,
a steady bifurcation of the zero solution of (\ref{e:modhopf}) occurs;
this bifurcation produces the phase-locked standing waves $PSW$ with
frequency $\omega_e/n$.  The amplitude $r=|v_1|=|v_2|$ of the $PSW$
satisfies
\be
|N|^2 r^4+2r^2 Re\left\{\nu^*N\right\} + |\nu|^2-\mu^2=0, \label{e:pswbr}
\ee
where $N\equiv\alpha+\beta$. If $Re\{\nu^*N\}\equiv \nu_r N_r + \nu_i
N_i>0$ the bifurcation to $PSW$ is forward, otherwise backward. Thus,
depending on the detuning $\nu_i$, the bifurcation to the $PSW$ can be
in either direction, independent of the bifurcation direction of the
standing waves in the unforced Hopf bifurcation. The saddle-node line
of the $PSW$, marked $SN$ in fig.\ref{f:pdsup}, is given by
\be
\mu^2=(Im\{\nu^* N\})^2/|N|^2, \label{e:pswsn}
\ee
for $Re\{\nu^* N\}<0$.

The linear stability analysis of the $PSW$ is simplified by the fact
that the perturbations $(u_1,u_2)$ of the steady solution
$(v_1,v_2)=(v,v)$ of (\ref{e:modhopf}) fall into two classes:
$u_1=u_2$ and $u_1=-u_2$. We focus first on the perturbations that
preserve the standing-wave character of the waves ($u_1=u_2$).  The
determinant and trace of the resulting $2\times 2$ stability matrix
are
\bea
Det=|\nu+2Nr^2|^2-|\nu|^2,\\
Tr=2(\nu_r+2N_r r^2),
\eea
where $r^2$ solves (\ref{e:pswbr}).  The determinant vanishes at the
saddle-node bifurcation (\ref{e:pswsn}). There is a secondary Hopf
bifurcation when the trace vanishes, provided the determinant is
positive. This bifurcation produces unlocked standing waves $SW$.  The
perturbations of the form $u_1=-u_2$ include the marginal translation
mode and a parity-breaking instability to traveling waves. The
parity-breaking instability occurs when the trace of the associated
Jacobian matrix vanishes, {\it i.e.} when
\be
Tr=2(\nu_r+2\alpha_r r^2)=0.\label{e:pswpb}
\ee
It is denoted by $PB$ in fig.\ref{f:pdsup}.  A typical bifurcation
diagram is presented in fig.\ref{f:pswbif}.  As long as the unforced
traveling waves are stable with respect to the standing waves
($\beta_r<0$) the secondary Hopf bifurcation of the $PSW$ to unlocked
standing waves is always preempted by the parity-breaking bifurcation
to traveling waves. In addition, modulated waves exhibiting three
different frequencies are possible; this follows from an analysis of
eq. (\ref{e:modhopf}) in the vicinity of the point where the Hopf
bifurcation and the steady bifurcation merge ($\mu^2=\nu_i^2,
\nu_r=0$)
\cite{rck88}. This codimension-two point corresponds to a
Takens-Bogdanov bifurcation with $O(2)$-symmetry, which was analyzed
by Dangelmayr and Knobloch \cite{dk87}.

Thus the stability of the phase-locked states ($PSRo$, $PSRe$) to
traveling disturbances ($TRo$, $TRe^{\perp,\parallel}$) and to unlocked
standing waves ($SRo$, $SRe$) is understood in some detail based on
previous studies. The competition between the different kinds of
phase-locked standing waves is the subject of the next section.

\newpage
\section{Competition between Phase-Locked Standing Rolls and
Standing Rectangles}

The relative stability of $PSRo$ and $PSRe$ is particularly relevant
to our understanding of the EHC experiments since both states are
observed. This stability question can be addressed by restricting
(\ref{e:agl3}) to the standing-cross-rolls subspace where
$(z_1,z_2,z_3,z_4)=(v_1,v_2,v_1,v_2)$. Note that this subspace
contains both the $PSRo$ ($v_1=0$ or $v_2=0$) and the $PSRe$
($v_1=v_2$) solutions. The amplitudes $v_1$ and $v_2$ satisfy
\bea
\dot{v}_1&=&\nu v_1+\mu v_1^{*}+(a+c) v_1|v_1|^2
+(b+d) v_1|v_2|^2+f v_1^*v_2^2,
\nonumber \\
\dot{v}_2&=&\nu v_2+\mu v_2^{*}+(b+d) v_2|v_1|^2
+(a+c) v_2|v_2|^2+f v_2^*v_1^2.
\label{e:pscr}
\eea
These equations differ from those describing the forced $O(2)$-Hopf
bifurcation (\ref{e:modhopf}); the modes with amplitudes $v_1$ and
$v_2$ are uncoupled at the linear level since they represent standing
waves with different orientations. Moreover, the nonlinear coupling
term with coefficient $f$ is absent in (\ref{e:modhopf}).

We focus first on the stability of the $PSRo$ solution, which has the
form $(v_1,v_2)=(u,0)$, to perturbations in the $(0,v_2)$-direction.
The resulting $2\times 2$ stability matrix $D_{PSRo}$ has the
following trace and determinant:
\bea
Tr(D_{PSRo}) &=& 2(\nu_r+(b_r+d_r)r_{PSRo}^2), \label{e:hopfpsr}\\
Det(D_{PSRo})&=& |\nu +(b+d)r_{PSRo}^2|^2-|\nu +(a+c-f
)r_{PSRo}^2|^2.\label{detpsr}
\eea
The amplitude $r_{PSRo}$ satisfies (\ref{e:pswbr}) with $N=a+c$.
Provided the determinant (\ref{detpsr}) is positive, there is a Hopf
bifurcation at $\nu_r=-(b_r+d_r)r_{PSRo}^2$; thus the presence of the
$PSRo$ in one direction leads to a shift of the primary Hopf
bifurcation in the other direction. A steady state bifurcation occurs
for
\be
\frac {r_{PSRo}^2}{2}
=\rho\equiv\frac {Re\left\{\nu^{*}(a+c-b-d-f)\right\}}{|b+d|^2-|a+c-f|^2}.
\label{e:defrho}
\ee
We will show that this bifurcation leads to phase-locked standing
cross rolls $PSCR$ which are also pertinent to the EHC experiments.
The $PSRo$ are stable in the $(0,v_2)$-direction as long as the trace
(\ref{e:hopfpsr}) is negative and
\be
2 Re\left\{\nu^{*}(a+c-b-d-f)\right\}
< \left(|b+d|^2-|a+c-f|^2 \right)r^2_{PSRo}.
\label{e:psrstab}
\ee
Note that $\rho$ depends on both the control parameters $\nu_r$ and
$\nu_i$.  Moreover, since the amplitude $r_{PSRo}$ can be varied
independent of $\nu$ by changing the forcing amplitude $\mu$,
eq.(\ref{e:psrstab})
reduces at onset to
\be
2 Re\left\{\nu^{*}(a+c-b-d-f)\right\} < 0. \label{e:PSRostabon}
\ee
Finally we note that the symmetry of the $PSRo$ solution results in a
doubling of the eigenvalues associated with perturbations at an
oblique angle to the rolls (see, for example, \cite{sk91}). Hence the
stability calculations of this section actually determine four of the
eight eigenvalues of the full stability matrix associated with
(\ref{e:agl3}) linearized about $PSRo$; the remaining four eigenvalues
were determined in the previous section.

We now examine the stability of the phase-locked standing rectangles
($PSRe$) in the standing-cross-rolls subspace. The $PSRe$ solution is
an equilibrium solution of (\ref{e:pscr}) of the form
$(v_1,v_2)=(u,u)$. We examine its stability with respect to
perturbations of the form $(w,-w)$ by substituting
$(v_1,v_2)=(u+w,u-w)$ in (\ref{e:pscr}). The resulting $2\times 2$
stability matrix $D_{PSRe}$ has the following trace and determinant:
\bea
Tr(D_{PSRe})&=&2\nu_r+4(a_r+c_r-f_r)r_{PSRe}^2, \label{e:hopfpsb}\\
Det(D_{PSRe})&=&|\nu +2(a+c-f)r_{PSRe}^2|^2-|\nu +2(b+d)r_{PSRe}^2|^2,
\eea
where $r_{PSRe}$ is a solution of (\ref{e:pswbr}) with $N=a+b+c+d+f$.
Note that
\be
Det\left(D_{PSRo}(r_{PSRo}^2)\right)=
-Det\left(D_{PSRe}(2r_{PSRe}^2)\right).\label{e:detsrsb}
\ee
Consequently a steady bifurcation to $PSCR$ occurs at
\begin{equation}
r_{PSRe}^2=\rho, \label{e:psbpscr}
\end{equation}
where $\rho$ is defined in (\ref{e:defrho}).  The coincidence of the
steady state bifurcations of the $PSRe-$ and the $PSRo-$branches, both
to $PSCR$ solutions, suggests that they may be connected by the {\it
same} branch of $PSCR$; this provides a mechanism for a continuous
transition between $PSRo$ and $PSRe$. We will show, in a particular
limit, that this possibility can occur.

The $PSCR$ are steady state solutions of (\ref{e:pscr}) with
$v_1v_2\ne 0$ and $|v_1|\ne|v_2|$.  We let $v_1=Re^{i\Phi}$ and
$v_2=re^{i\phi}$, and  eliminate the phases from (\ref{e:pscr}) to
obtain (for $r \ne 0$)
\begin{equation}
\mu^2=|\nu +(a+c) r^2+(b+d) R^2+fR^2\frac {
\nu +(a+c - f )r^2+(b+d)R^2}{\nu +(b+d)r^2+(a+c-f
)R^2}|^2. \label{e:pscr56}
\end{equation}
Moreover, if $r$ and $R$ are both non-zero, then
\be
|\nu+(b+d) R^2+(a+c-f)r^2|^2=|\nu+(a+c-f)R^2+(b+d)r^2|^2.
\ee
This has a solution
$r^2=R^2$,
which corresponds to the $PSRe$, and also a solution
\be
r^2+R^2=2\rho. \label{e:pscr510}
\ee
Substituting $r^2=2\rho-R^2$ in (\ref{e:pscr56}) leads to a quartic
equation for $R^2$:
\be
C_8 R^8+C_6 R^6+C_4 R^4 + C_2 R^2 +C_0 =0. \label{e:biquartic}
\ee
The coefficients $C_0$, $C_2$ and $C_8$ are given by
\bea
C_0&=&|\nu+2(b+d)\rho |^2 \left( |\nu + 2(a+c) \rho |^2
-\mu^2\right),\label{e:defC0}\\
C_2&=&2Re \left\{ \left(\nu+2(b+d) \rho \right) \left(h^*-f^*\right)
\left( 2 \rho (\nu+2(a+c)\rho) (h^*+f^*)
-\mu^2\right) \right\},\\
\label{e:defC1}
C_8&=&|h^2-f^2|^2,
\label{e:defC8}
\eea
where $h\equiv a+c-b-d$. We use the reflection symmetry that
interchanges $R^2=|v_1|^2$ and $r^2=|v_2|^2$ to simplify this
equation.  In particular, we observe that for each solution $R_1^2$
there must also exist a solution $R_2^2=2\rho-R_1^2$ ({\it cf}.
(\ref{e:pscr510})).  We let
\be
u\equiv R^2(2\rho-R^2)
\ee
and write the solutions to (\ref{e:biquartic}) as
\be
u=\frac{-C_2 \pm \sqrt{C_2^2-16 C_0 C_8\rho^2}}{4\rho C_8}\ ,
\label{e:ueq}
\ee
which yields the symmetry-related pair of solutions
\be
R^2=\rho\pm\sqrt{\rho^2-u}. \label{e:solnr2}
\ee
Note that both $u$ and $R^2$ must be positive for $PSCR$ to exist.
{}From (\ref{e:ueq}), it follows that there are at most {\it two}
distinct $PSCR$-solutions of (\ref{e:pscr}); these annihilate in a
saddle-node bifurcation when $C_2^2=16C_0 C_8\rho^2$. We note that the
$PSCR$-solution merges with the $PSRo$ when $u=0$ and with $PSRe$ when
$u=\rho^2$.
We now consider two limiting cases in which we can show explicitly
that the $PSCR$ arise as a secondary solution branch that connects the
$PSRo$- and $PSRe$-solutions.

The phase-locked standing waves arise through a steady bifurcation at
$|\nu|^2=\mu^2$, whenever $\nu_r<0$, {\it i.e.}, for values of $\nu_r$
below the unforced Hopf bifurcation point. We can use center manifold
reduction
\cite{gh} to reduce the complex equations (\ref{e:pscr}) to two real
equations for the standing-wave amplitudes \cite{r90a}. For
$|\nu|^2=\mu^2$, $\nu_r<0$, the neutral eigenspace of (\ref{e:pscr})
linearized about the origin is spanned by $v_j=\eta_1\equiv
i(\mu+\nu^*)$, $j=1,2$; the damped eigendirections associated with the
double eigenvalue $2\nu_r$ are given by $v_j=\eta_2\equiv i(\mu-\nu)$.
We introduce the slow time $\tau=\epsilon^2 T$ and let $v_j=\epsilon
\eta_1 A_{j}+\epsilon^3 (\eta_1 A_{j3}+\eta_2 B_{j3}) + h.o.t$,
$\mu=|\nu|+\epsilon^2\mu_2$. At cubic order in $\epsilon$ we obtain
\bea
\frac{dA_1}{d\tau}=
\sigma A_1+\tilde{\alpha}A_1^3+\tilde{\beta}A_1A_2^2, \nonumber\\
\frac{dA_2}{d\tau}=
\sigma A_2+\tilde{\alpha}A_2^3+\tilde{\beta}A_2A_1^2,\label{e:sthopf}
\eea
where
\be
\sigma=-\frac{|\nu| \mu_2}{\nu_r},
\quad \tilde{\alpha}=\frac{|\eta_1|^2}{\nu_r}Re\{\nu^*(a+c)\},
\quad \tilde{\beta}= \frac{|\eta_1|^2}{\nu_r}
Re\{\nu^*(b+d+f)\}. \label{e:alphabeta}
\ee
Equations of the form (\ref{e:sthopf}) also apply to
systems with large damping whenever sufficiently large forcing is
applied. In particular, these equations describe the oblique
standing-wave components that have the correct temporal phase, relative
to the periodic driver, to be excited by the forcing. In contrast,
eqs.(\ref{e:agl3}) become invalid for large damping.
Note that in general
the amplitudes in (\ref{e:sthopf}) are complex due to translation symmetry;
the specific choice for the standing-cross-rolls subspace ($z_1 \equiv z_3, z_2
\equiv z_4$)
 that leads to eq.(\ref{e:pscr})
breaks that symmetry.

Equations of the same form as (\ref{e:sthopf}) arise in the analysis
of the $O(2)$-Hopf bifurcation problem and have been studied
extensively \cite{gss88}.  Generically, the only steady solutions of
(\ref{e:sthopf}) have either equal amplitudes ($A_1=A_2$), or one of
the amplitudes vanishes. Thus, at onset one has only $PSRo$ (one
vanishing amplitude) and $PSRe$ ($A_1=A_2$). Only in special
degenerate cases do we expect more complicated small amplitude
equilibrium solutions near onset. In particular, if $\tilde{\alpha}
\approx
\tilde{\beta}$, then higher order terms must be included in the normal
form (\ref{e:sthopf}) and there can exist a small-amplitude mixed-mode
equilibrium solution with two different non-vanishing amplitudes
\cite{gr87,ck88}.  The mixed-mode solution corresponds to the $PSCR$.
Note that $\rho
\propto (\tilde\alpha-\tilde\beta)$ so that $\rho$ is small in this
degenerate situation. This is consistent with (\ref{e:defrho}) and
(\ref{e:psbpscr}) which show that the secondary bifurcations to $PSCR$
occur when the amplitudes $r^2_{PSRo}$ and $r^2_{PSRe}$ are of order
$\rho$, respectively.  In the two-dimensional control parameter space
($\mu_2, \tilde{\alpha}-\tilde{\beta}$) the regime in which the $PSCR$
exist is delimited by the lines at which the $PSRo$ and $PSRe$ become
unstable to $PSCR$. The stability of the $PSCR$ solution depends on
higher order terms which were neglected in deriving the amplitude
equations (\ref{e:agl3}) \cite{gr87,ck88}.

The center manifold equations (\ref{e:sthopf}) lose their validity
when $\nu_r\rightarrow 0$ since the eigendirections with eigenvalue
$2\nu_r$ can no longer be eliminated adiabatically; these standing
wave modes become independent dynamical variables.  For nonzero
detuning $\nu_i$ this regime can be analyzed by considering the
Takens-Bogdanov double-zero point ($\nu_r=0,\mu=\mu_0\equiv |\nu_i|$)
at which the Hopf frequency goes to zero.  Near this point
simplification is not achieved by a reduction in the number of degrees
of freedom.  Instead, the form of the linear operator allows certain
nonlinear terms to be removed \cite{gh,etbci}. We begin by expanding
the amplitudes along the two directions $\zeta_1=(s+i)\mu_0$ and
$\zeta_2=1$ in the complex plane, where $s\equiv sgn(\nu_i)$
distinguishes between the cases of positive and negative detuning. Let
\be
v_j= n(\zeta_1 \hat{A}_j + \zeta_2 \hat{B}_j), \quad j=1,2, \quad  0< n \ll 1.
\label{e:tbexpa}
\ee
The new dynamical variables $\hat A_j$ and $\hat
B_j$ are real. At the double-zero point the linear part of
(\ref{e:pscr}) is
\begin{mathletters}
\bea
\dot{\hat{A}_j}&=&s\hat{B}_j, \label{e:tb0A}\\
\dot{\hat{B}_j}&=&0.\label{e:tb0B}
\eea
\end{mathletters}
Since $s\ne 0$, we can perform a near-identity nonlinear coordinate
transformation that removes most of the cubic terms in the normal
form. In particular, we let
\begin{mathletters}
\bea
A_1&=&\hat A_1+n^2(c_1 \hat A_1^2+c_2\hat B_1\hat A_1+c_3\hat A_2^2+c_4
\hat B_2\hat A_2+c_5\hat B_2^2)\hat A_1\nonumber\\
&&+ n^2(c_6 \hat B_1^2+c_7\hat B_1\hat A_1+c_8\hat A_2^2+c_9
\hat B_2\hat A_2+c_{10}\hat B_2^2)\hat B_1,\label{e:nfca1}\\
B_1&=&\hat B_1+n^2(d_1 \hat A_1^2+d_2\hat B_1\hat A_1+d_3\hat A_2^2+d_4
\hat B_2\hat A_2+d_5\hat B_2^2)\hat A_1\nonumber\\
&&+ n^2(d_6 \hat B_1^2+d_7\hat B_1\hat A_1+d_8\hat A_2^2+d_9
\hat B_2\hat  A_2+d_{10}\hat B_2^2)\hat B_1,\label{e:nfcb1}
\eea
\end{mathletters}
with $A_2$ and $B_2$ defined by interchanging the $1$ and $2$
subscripts in (\ref{e:nfca1}) and (\ref{e:nfcb1}), respectively. We
choose the coefficients $c_j$ and $d_j$ so as to remove the cubic
terms in the $\dot{\hat{A}_j}$ equations and four of the ten cubic
terms in the $\dot{\hat{B}_j}$ equations. Moreover, in order to
express the unfolding parameters of the resulting equations in terms
of the original control parameters, we expand $\nu$ and $\mu$ as
\be
\mu=\mu_0+\epsilon \mu_1,
\quad \nu= is\mu_0+\epsilon(\nu_{r1}+i\nu_{i1}), \quad
\label{e:tbexpp}
\ee
where $0<\epsilon\ll 1$.
The transformed equations are
\begin{mathletters}
\bea
\dot{A}_1&=&s B_1 + \epsilon (\nu_{r1}+s\nu_{i1}-\mu_1)A_1+
\epsilon \nu_{i1}B_1/\mu_0 +
{\cal O}(\epsilon n^2,n^4) \label{e:tbnormA},\\
\dot{B}_1&=&\epsilon (\nu_{r1}-s \nu_{i1} + \mu_1 ) B_1 - 2 \epsilon
\mu_0 (\nu_{i1} - s \mu_1) A_1  \nonumber \\
& &+4n^2 \{
- \mu_0^3 (a_i+c_i) A_1^3 + 2 \mu_0^2 (a_r+c_r) B_1 A_1^2
-  \mu_0^3 (b_i+d_i +f_i) A_1 A_2^2  \nonumber \\
& &+  \mu_0^2 (b_r+d_r+2 f_r) B_2 A_1 A_2
 +  \mu_0^2 (b_r+d_r) B_1 A_2^2 +
 \mu_0 (b_i+d_i-2f_i) B_2 B_1 A_2 \} \nonumber \\
& &
+{\cal O}(\epsilon n^2,n^4), \label{e:tbnormB}
\eea
\end{mathletters}
with similar equations for $A_2$ and $B_2$. We focus on the stationary
solutions of (\ref{e:tbnormA},\ref{e:tbnormB}), which correspond to
the phase-locked standing waves observed in the experiments. In this
case, (\ref{e:tbnormA}) yields $B_j={\cal O}(\epsilon, n^4)$, $j=1,2$,
which takes all terms involving $B_j$ in (\ref{e:tbnormB}) to higher
order.  We obtain the following algebraic equations for the $A_j$:
\bea
0&=&-2\mu_0\bigl[\epsilon (\nu_{i1} - s \mu_1) +2 n^2 \mu_0^2\left(
(a_i+c_i)A_1^2+(b_i+d_i+f_i)  A_2^2    \right)\bigr]A_1
+ {\cal O}(\epsilon^2,\epsilon n^2,n^4), \nonumber \label{e:tb01}\\
0&=& -2\mu_0\bigl[ \epsilon  (\nu_{i1} - s \mu_1) +2  n^2 \mu_0^2 \left(
(a_i+c_i) A_2^2  +(b_i+d_i+f_i) A_1^2   \right)\bigr]A_2
+ {\cal O}(\epsilon^2,\epsilon n^2,n^4). \label{e:tb02}
\eea
Note that the equations for the equilibrium solutions of
(\ref{e:sthopf}) are of the same form as (\ref{e:tb02}). Thus, as
before, $PSRo$ and $PSRe$ are the only phase-locked standing waves in
the vicinity of the Takens-Bogdanov point, unless $a_i+c_i\approx
b_i+d_i+f_i$, in which case higher order terms must be retained.  In
this case, it is known that the existence region of $PSCR$ lies
between steady state instability lines of $PSRo$ and $PSRe$
\cite{ck88}. Again this degenerate case corresponds to $\rho\approx
0$. (Recall that $\nu_r\approx 0$ near the Takens-Bogdanov point.)

We can combine the above results to obtain the following picture. The
steady state instability at $\mu=|\nu|$ produces both $PSRo$ and
$PSRe$, with at most one of the two states being stable.  This follows
from (\ref{e:detsrsb}); a positive determinant is necessary for
stability and for $r^2_{PSRo,PSRe}
\rightarrow 0$ the determinants of $D_{PSRo}$ and $D_{PSRe}$  have
opposite signs. (Recall that $r^2$ can be decreased by varying the
control parameter $\mu$, while holding $\nu$ fixed.)  In addition, for
$\nu_r$ sufficiently negative, {\it i.e.} below the onset of unforced
traveling waves, the trace of $D_{PSRo}$ (\ref{e:hopfpsr}) and
$D_{PSRe}$ (\ref{e:hopfpsb}) are both negative. For $\rho$
sufficiently small, the steady bifurcations from $PSRo$ and $PSRe$ to
$PSCR$ can occur before the Hopf bifurcations associated with
$Tr(D_{PSRo})=0$ and $Tr(D_{PSRe})=0$. The difference in the steady
state bifurcation values for each of the solutions is given by
\bea
\Delta\mu^2 &\equiv&\mu_{PSRo\rightarrow PSCR}^2-\mu^2_{PSRe\rightarrow
PSCR}\nonumber \\
&=&\rho^2\left(2(|b+d|^2-|a+c-f|^2)-(|a+c+b+d+f|^2-4|a+c|^2)\right).\label{e:deltamu}
\eea
The two bifurcations coincide for $\rho=0$ in which case they lie on
the neutral curve $\mu^2=|\nu |^2$. For $\rho<0$ there is no
bifurcation to $PSCR$. For $\rho >0$ there are two main cases
distinguished by the sign of $|b+d|^2-|a+c-f|^2$ ({\it cf.}
(\ref{e:psrstab})). In these two cases, the conditions for stability
of $PSRo$ and $PSRe$, respectively, to $PSCR$ are given by
\begin{mathletters}
\bea
\mbox{I\ }&(|b+d|^2-|a+c-f|^2>0):\quad & PSRo:\ r_{PSRo}^2>2\rho, \quad
PSRe:\ r_{PSRe}^2<\rho,\\
\mbox{II\ }&(|b+d|^2-|a+c-f|^2<0):\quad & PSRo:\ r_{PSRo}^2<2\rho, \quad
PSRe:\ r_{PSRe}^2>\rho.
\label{cases}
\eea
\end{mathletters}

Figures \ref{f:psrpsba} and \ref{f:psrpsbb} present possible
bifurcation diagrams when $\rho>0$, and when both $PSRo$ and $PSRe$
bifurcate supercritically.  In case I, the $PSRe$ are stable at onset,
but become unstable to $PSCR$ at larger amplitude ({\it i.e.} larger
values of $\mu^2$). For $\Delta\mu^2>0$, the $PSCR$ merge with the
$PSRo$ for even larger $\mu^2$ and stabilize them in turn. Our
analysis does not determine whether the pitchfork bifurcation to the
$PSCR$ is supercritical or subcritical.  In the simplest scenario, the
bifurcation is supercritical when $\Delta\mu^2>0$ and the $PSCR$
solution is stable in the $SCR$ subspace. For $\Delta\mu^2<0$, the
transition from $PSRe$ to $PSRo$ is hysteretic and (in the simplest
case) the $PSCR$ are unstable.  In case II the $PSRo$ and $PSRe$
interchange their roles and stable $PSCR$ are expected for
$\Delta\mu^2<0$. If the $PSRo$ or $PSRe$ bifurcate subcritically from
the basic state more complicated bifurcation diagrams are possible.

We emphasize that $\rho$ depends on the control parameter $\nu$.
Within the present framework, one therefore expects that it may be
possible to choose the detuning $\nu_i$ such that $\rho$ is in a
suitable range to observe the transition to $PSCR$ below the onset of
unforced waves ($\nu_r<0$) and before other instabilities ({\it e.g.},
parity breaking) set in.  In particular, by choosing $\nu_i$ such that
$\rho$ is small and positive it may be possible to observe the
transition between $PSRo$ and $PSRe$ by increasing $\mu$. Whether this
transition is made via a stable or unstable branch of $PSCR$ solutions
will depend on the nonlinear coefficients (see (\ref{e:deltamu})).
Similarly, if both $PSRo$ and $PSRe$ bifurcate supercritically and
$\nu_r$ is sufficiently negative, then $\rho$ determines whether
$PSRo$ or $PSRe$ are stable at onset. In particular, $PSRo$ arise
stably at onset if $\rho <0$ in case I and if $\rho >0$ in case II
(see (\ref{e:defrho}) and (\ref{e:PSRostabon})).  As will be discussed
below, the range over which the detuning and therefore $\rho$ can be
varied effectively is, however, limited in spatially extended systems.

\section{Temporal Forcing in the Vicinity of a Lifshitz Point }

Experiments on the liquid-crystal system have identified a parameter
regime where rolls travel along the axis of anisotropy as well as a
parameter regime where (disordered) waves travel at an oblique angle
to it. This is reminiscent of the situation of {\it steady} normal and
oblique rolls for which the transition between the two regimes is
usually continuous and the angle of obliqueness,
$\gamma=\tan^{-1}(p/q)$ ({\it cf.} (\ref{e:vz})), goes to zero at the
Lifshitz point in parameter space
\cite{zk85,rwtrs89}. In the present case of Hopf bifurcation to
traveling rolls, the disordered character of the unforced waves has
made experimental confirmation of the existence of such a Lifshitz
point problematic; no such experimental verification exists to date.
Nevertheless, such a point is of special interest since it provides a
natural mechanism for a transition between the normal and oblique
regimes. At the Lifshitz point, the critical modes with amplitudes
$z_1$ and $z_4$ in (\ref{e:vz}) become indistinguishable as $\pm
p\rightarrow 0$, as do the critical modes with amplitudes $z_2$,
$z_3$; the description in terms of four ordinary differential
equations is no longer appropriate. Instead, the behavior in the
vicinity of the Lifshitz point is modeled by a generalized
Ginzburg-Landau equation in which the small wave number $p$ is
incorporated into a slow variable in the $y$-direction. This
description of the Hopf bifurcation in the vicinity of the Lifshitz
point is analogous to the model of the Lifshitz point in the steady
state case \cite{pk86}. The latter leads to a single partial
differential equation (in two space dimensions) with real
coefficients, whereas in the Hopf case the system is described by two
coupled complex partial differential equations.

For a Hopf bifurcation with translation and reflection symmetry two
complex amplitudes are required to model the behavior of both the
left- and right-traveling waves.  We replace (\ref{e:vz}) by
\be
{V_z(x,y,t)}=\epsilon^2A(x,y,t)e^{i(\omega_e\bar {t}/n+\bar{q}\bar
{x})}+\epsilon^ 2B(x,y,t)e^{i(\omega_e\bar {t}/n-\bar{q}\bar
{x})}+h.o.t.+c.c.,
\ee
$0<\epsilon\ll 1$, where $x=\epsilon^2\bar {x}$, $y=\epsilon\bar {y}$
and $t=\epsilon^ 4\bar {t}$ are slow variables.  The spatial
symmetries of the system result in the following equivariance
properties of the generalized complex Ginzburg-Landau equations:
\begin{mathletters}
\bea
&\bar {x}\rightarrow\bar {x}+\theta /\bar{q}:\quad &
(A,B,\partial_x,\partial_y)\rightarrow
(Ae^{i\theta},Be^{-i\theta},\partial_x,
\partial_y),\\ &\bar {x}\rightarrow -\bar {x}:&
(A,B,\partial_x,\partial_y)\rightarrow (B,A,-\partial_x,
\partial_y),\\ &\bar
{y}\rightarrow -\bar {y}:& (A,B,\partial_x,\partial_y)\rightarrow
(A,B,\partial_x,
-\partial_y).
\label{e:glsym}
\eea
\end{mathletters}
One then obtains the following equations (through order $\epsilon^6$)
\bea
\partial_tA-v\partial_xA=d\partial_x^2A+s\partial_y^2A+
g\partial_x\partial_y^2A+h\partial_y^4A+\nu A+\mu B^{*}\nonumber \\
+\tilde{c}A(|A|^2+|B|^2)+\tilde{g}A|B|^2, \nonumber \\
\partial_tB+v\partial_xB=d\partial_x^2B+s\partial_y^2B-
g\partial_x\partial_y^2B+h\partial_y^4B+\nu B+\mu A^{*}\nonumber \\
+\tilde{c}B(|A|^2+|B|^2)+\tilde{g}B|A|^2,\label{e:lgl}
\eea
where, without loss of generality, we assume that the group velocity
$v$ and the (scaled) forcing amplitude $\mu$ are real; all other
coefficients are complex.  The Lifshitz equations (\ref{e:lgl})
without the temporal forcing ({\it i.e.} $\mu=0$) were introduced in
\cite{z91} and partially analyzed in \cite{srk92}.  In the unforced case the
Lifshitz point is given by $s_r=\nu_r=0$, with the unscaled deviation
from the Lifshitz point given by ($\epsilon^2 s_r$, $\epsilon^4
\nu_r$).  As before we assume the temporal driving is small and
therefore keep only the lowest order terms $\mu B^{*}$ and $\mu A^{*}$
that destroy the phase-shift symmetry $(A,B)\to e^{i\phi}(A,B)$.
Specifically, the unscaled forcing is $\bar \mu=\epsilon^4\mu$.
Similarly, the complex coefficient $\nu$ has been scaled by
$\epsilon^4$, {\it i.e.} $\bar\nu=\epsilon^4\nu$. It should be
emphasized that all terms in (\ref{e:lgl}) are not generally ${\cal
O}(1)$ in $\epsilon$. In particular, if the (unscaled) group velocity
($\bar{v}$) and the unscaled imaginary part of the $s$ ($\bar{s}_i$)
are ${\cal O}(1)$, then $v$ and $s_i$ are ${\cal O}(1/\epsilon^2)$.
Formally, we assume that $\bar v$ and $\bar s_i$ are ${\cal
O}(\epsilon^2)$ so that $\bar{v}=\epsilon^2 v$, $\bar{s}=\epsilon^2
s$. In an asymptotic analysis of the general case, where $\bar v$ and
$\bar s_i$ are ${\cal O}(1)$, one would introduce a faster time
$t_1=\epsilon^2\bar {t}$ and first solve the ${\cal O}(\epsilon^4)$
equations before introducing those at ${\cal O}(\epsilon^6)$
\cite{kd90,mv92}. We expect to capture part of this general
case by assuming $v$ and $s_i$ are large ({\it i.e.} ${\cal
O}(1/\epsilon^2)$) in the final results.

Without temporal forcing, the growth rate $\sigma$ of modes $A, B
\propto e^{\pm i(qx+py)}$ is given by $\sigma
=\nu_r-d_rq^2-s_rp^2+g_iqp^2+h_rp^4$. Thus, for $s_r\ge 0$, a Hopf
bifurcation to normal traveling rolls occurs at
\be
 \nu_r=0, \  q_c=0, \ p_c=0.
\label{hopfnormal}
\ee
For $s_r<0$ the bifurcation to oblique traveling waves is given by
\bea
\nu_r=\frac{1}{2}s_rp_c^2, \ q_c=\frac{g_is_r}{4h_rd_r+g_i^2},
\ p_c^2=\frac{2d_rs_r}{4h_rd_r+g_i^2}> 0.
\label{hopfoblique}
\eea
Note that stability of the trivial state with respect to large wave
numbers $q$ and $p$ requires that $d_r >0, h_r<0$. Moreover, we assume
$4h_rd_r+g_i^2<0$ so that the instability to oblique waves occurs when
$s_r<0$, with $\nu_r<0$ in (\ref{hopfoblique}).

With resonant temporal forcing, the location of the Hopf bifurcation
to normal, as well as to oblique, waves remains unchanged for small
$\mu$, {\it i.e.} for $0<\mu<|\nu_{eff}(q_c,p_c)|$, where
\be
\nu_{eff}(q,p)=\nu+iqv-dq^2-sp^2-igqp^2+hp^4.
\label{e:nueff}
\ee
The frequency on the Hopf bifurcation surface is given by
$\omega=\sqrt{|\nu_{eff}|^2-\mu^2}$. Recall that this frequency is the
difference between half the driving frequency (for $n=2$) and the
frequency of the critical mode.  Note that the frequency $\omega$
decreases as $\mu$ increases, until it goes to zero for
$\mu^2=\mu^2_0(q,p)\equiv|\nu_{eff}|^2$. At such points, the Hopf
bifurcation is replaced by a steady bifurcation of the trivial
solution of the amplitude equations (\ref{e:lgl}). This steady
bifurcation leads to (standing) waves which are phase-locked to the
forcing.

To get some insight into the nonlinear solutions of (\ref{e:lgl}) we
consider two limiting cases. Since the phase-locked waves arise
through a bifurcation involving a real eigenvalue, the coupled complex
equations (\ref{e:lgl}) can be reduced to a single equation
sufficiently close to the neutral surface \cite{r90a}. Depending on
the kind of extremum $(q_c,p_c)$ chosen as expansion point, different
equations are obtained. Here we concentrate on a situation that
eventually leads to an equation of the same form as for the steady
bifurcation at a Lifshitz point \cite{pk86}. Another limiting case is
obtained by considering oblique waves within (\ref{e:lgl}), {\it i.e.}
assuming $|p| >0$ for small $\nu$ and $\mu$. This allows us to derive the
amplitude equations (\ref{e:agl3}) from the Lifshitz equations.

 We
first describe the reduction to a single real Lifshitz equation.
The condition for a Lifshitz point for the standing waves on the normal-roll
branch is
\be
\partial_p^2\mu^2(q_c,p_c=0)=0
\ee
where $q_c$ is determined by $\partial_q \mu^2(q_c,p_c=0)=0$. These two
conditions determine
the critical wavenumber $q=q_c$ and the Lifshitz point in parameter space
$s_r=s_r^{LP}$, respectively:

\bea
-2(\nu_r-d_rq_c^2)d_rq_c+(vq_c+\nu_i-d_iq_c^2)(v-2d_iq_c)=0,\\
s_r^{LP}=\frac {(\nu_i+vq_c-d_iq_c^2)(s_i+g_rq_c)+g_iq_c(d_rq_c^2-\nu_r)}{d_r
q_c^2-\nu_r}.
\eea
Note that the presence of the forcing has shifted the Lifshitz point
for the standing waves to a non-zero value of $s_r$. This is in
contrast to the unforced case where the Lifshitz point occurs at
$s_r=0$ with $q_c=p_c=0$. For large values of $v$ and $s_i$, of order
$\epsilon^{-2}$, this leads to
\be
q_c=-\frac {\nu_i}v\left(1+\frac {2\nu_rd_r-\nu_id_i}{v^2}\right)+{\cal O}
(\epsilon^8) \label{e:qclif}
\ee
and a shift of the Lifshitz point to
\be
s_r^{LP}=-\frac 1{v}\left(\nu_ig_i-2d_r\frac {\nu_i s_i}{v}\right)
+{\cal O}(\epsilon^{6}).
\ee
Thus, in this limit of large group velocity $v$ in (\ref{e:lgl}), the
steady state neutral surface $\mu^2=\mu_0^2$ has only a single minimum
with $p=0$; this occurs at $\mu^2=\mu^2_c\equiv\mu_0^2(q_c,p_c=0)$,
where $q_c$ is given by (\ref{e:qclif}). (In the special case that $v$
is not large, this need not be the case as will be discussed below.)

To derive the reduced Lifshitz equation from eq.(\ref{e:lgl}), we
introduce super-slow space and time scales and expand $\mu$ about its
critical value $\mu_c=|\nu_{eff}(q_c,p=0,s_r^{LP})|$, $s_r$ about
its critical value $s_r^{LP}$, while holding $\nu$ fixed ($\nu_r<0$).
In particular, we let
\be
\mu=\mu_c+\delta^2\mu_2, \quad s_r = s_r^{LP} + \delta s_{r1},
\quad X=\delta x,
\quad Y=\delta^{1/2}y,\quad T=\delta^2t,
\ee
where $0<\delta\ll 1$. We use the two eigenvectors
$(\mu,-\nu_{eff})^t$ and $(\mu,\nu_{eff}^*)^t$ of the linear operator
evaluated on the neutral surface $\mu^2=|\nu_{eff}|^2$, and expand the
amplitudes $A$ and $B$ as follows:
\be
\left(\begin{array}{c}
A\\
B^{*}\end{array}
\right)=\delta\left(\begin{array}{c}
\mu\\
-\nu_{eff}\end{array}
\right)A_1e^{iq_cx}+\delta^2\left\{\left(\begin{array}{c}
\mu\\
-\nu_{eff}\end{array}
\right)A_2e^{iq_cx}+\left(\begin{array}{c}
\mu\\
\nu_{eff}^{*}\end{array}
\right)B_2e^{iq_cx}\right\}+{\cal O}(\delta^3).
\ee
At ${\cal O}(\delta^2)$ we obtain
\begin{equation}
B_2=-\frac 1{2\nu_{eff,r}}\left\{(v+2iq_cd)\partial_X+(s^{
LP}+iq_cg)\partial_Y^2\right\}A_1,\label{defb2}
\end{equation}
where $\nu_{eff,r}\equiv Re\{\nu_{eff}\}$ and $\nu_{eff}$ is evaluated at
$(q,p)=(q_c,0)$, and $s^{LP}\equiv s_r^{LP}+is_i$. Projecting the ${\cal
O}(\delta^3$)-equations onto the left-zero
eigenvector $(\mu,-\nu_{eff})$ the Lifshitz equation,
\begin{equation}
\partial_TA_1=(\kappa_1\partial_X^2+\kappa_2 \partial_Y^2+
i\kappa_3\partial_X\partial_Y^2+\kappa_4\partial_Y^4)A_1+\Sigma A_1+\Gamma
A_1|A_1|^2, \label{e:rlgl}
\end{equation}
is obtained, where the coefficients $\kappa_j$ ($j=1,...,4$), $\Sigma$
and $\Gamma$ are all real. We find that
\bea
\Sigma =-\frac{\mu_c \mu_2}{\nu_{eff,r}}, \quad
\Gamma =
\frac{\mu_c^2}{\nu_{eff,r}}Re\left\{\nu_{eff}^*(2\tilde{c}+\tilde{g})\right\}.
\label{e:rlglcoeff}
\eea
The coefficients $\kappa_i$ are given by the appropriate derivatives,
with respect to the wave numbers $q$ and $p$,
of the growth rate $\sigma = \nu_{eff,r} +
\sqrt{\mu^2-\nu_{eff,i}^2}$,
{\it e.g.} $\kappa_1=\partial_q^2 \sigma (q_c,p_c=0)$. Note that the
direction of bifurcation of the spatially uniform, (nontrivial) steady
solution of (\ref{e:rlgl}) depends on $\nu_{eff}$, as well as the
original nonlinear coefficients $\tilde c$ and $\tilde g$.

As mentioned before, eq.(\ref{e:rlgl}) has been studied in the case where the
bifurcation to the steady uniform solution is supercritical ({\it
i.e.} in the case where $\Gamma/\Sigma <0$)
\cite{pk86,bkkpwz91}. These investigations determined that
normal and oblique rolls can be stable in two-dimensional wave-number
regions, where the shape of these regions depends on the parameter
values.  In the present context, such solutions correspond to normal
phase-locked standing rolls ($NPSRo$) and oblique phase-locked
standing rolls ($OPSRo$). In the oblique-roll regime stable zig-zags
arise which connect $OPSRo$ of opposite angles with respect to the
director. In addition, undulated solutions can also be stable in small
islands in wave-number space \cite{pk86,bkkpwz91}.
In the subcritical case, suitable
nonlinear gradient terms in $X$- as well as $Y$-direction would have
to be included when going to fifth order.  They would allow the
direction of bifurcation of $NPSRo$ and of $OPSRo$ to differ from each
other.

The divergence of $B_2$ as $\nu_{eff,r}\rightarrow 0$ in
eq.(\ref{defb2}) shows that the reduced Lifshitz equation
(\ref{e:rlgl}) is no longer valid in this limit, {\it i.e.} as the
Hopf bifurcation point is approached. At the Hopf bifurcation, and
also at the parity-breaking bifurcations from standing structures to
traveling structures the full complex Lifshitz equations (\ref{e:lgl})
need to be used. This is not pursued here.

Additional information about possible phase-locked wave patterns is
obtained by examining the shape of the neutral surface
$\mu^2=|\nu_{eff}|^2$ in more detail, where $\nu_{eff}$ is given by
eq.(\ref{e:nueff}).  For example, an earlier study by Riecke, which
pertains to (normal) standing rolls in one spatial dimension, found
that the neutral curve need not be convex \cite{r90a,r90b}. In fact,
it can have two minima, in which case the band of stable wave numbers
splits into two disjoint regions, one centered at each of the two
minima. This has been investigated, in the one-dimensional case, using
the appropriate Ginzburg-Landau equations
\cite{r90a,rr93}, as well as phase equations \cite{r90a,rr93,bd90}.
Under the conditions that there are two minima of the neutral curve,
it was found that wave-number gradients need not decay as would
generally be expected if there was just a single minimum. Instead,
stable domain walls can arise which separate regions with different
wave numbers $q$.  In the present, two-dimensional study, one may
expect that multiple minima with differing values of $p$ can also be
obtained in the phase-locked standing wave regime. We investigate this
possibility by examining the form of the neutral surface, associated
with the generalized Ginzburg-Landau equations (\ref{e:lgl}), in a
special limiting case.

In the one-dimensional case it was shown that for multiple minima of
the neutral curve to occur the group velocity must be small
\cite{r90b} (see also equation (\ref{e:qclif})).  Thus, to demonstrate the
phenomena we choose $v=0$ as well as $g=0$. (The minima will then
persist for $v$ and $g$ sufficiently close to $0$.) Figs.
\ref{f:ns1} and \ref{f:ns2} show the neutral surface
$\mu^2=|\nu_{eff}(q,p)|^2$ for $\nu_r=-0.2$, $d=1+i$, $s=-1+0.585i$,
and $h=-4.27+2.5i$ for two different values of $\nu_i$. When
$\nu_i=-0.34$ (fig. \ref{f:ns1}), there is a minimum at $p_c=0$,
$q_c=0$, corresponding to normal $PSRo$, as well as a minimum (of
roughly the same depth) at $p=\pm p_c \approx 0.45$. Thus, in this
case one expects a competition between these three kinds of waves
which could lead to zig-zag patterns consisting of domains with
$p=p_c$ and $p=0$, respectively.  In contrast to the zig-zags
associated with the reduced Lifshitz equation (\ref{e:rlgl})
\cite{pk86}, one of the two zig-zag directions here is given by the
normal direction. Changing the frequency to $\nu_i=-0.3$ shifts the
minimum at $p_c=0$ to a non-zero value (fig. {\ref{f:ns2}). Now there
are four minima and one expects, for example, zig-zags consisting of
oblique waves with two different positive values of $p$. In the case
that $d_i<0$, the extremum at $q=0$ can become a maximum and two new
minima (in the $q$-direction) can arise at $q=\pm q_1$. (The
reflection symmetry that forces symmetry-related pairs $\pm q_1$, as
well as an extremum at $q=0$, is due to the special choice $v=0$ and
$g=0$.) Two cases with $d_i<0$ are shown in figs.
\ref{f:ns3} and \ref{f:ns4}.  For $d_i=-0.5$ (fig. \ref{f:ns3}) the
absolute minima are at $q=0$ and $p=\pm p_{1,2}$, whereas for
$d_i=-0.7$ (fig. \ref{f:ns4}) they are at $p=\pm p_1$ and $q=\pm q_1$.
Between these two situations the minima of the neutral surface almost
degenerate into a circular curve around the local maximum at $q=0$,
$p=\pm 0.35$. (Recall that in the physical system $q=0$ corresponds to
a non-zero wave number $\bar{q}$.) Such a neutral surface, with a
minimizing closed curve centered around a finite wave number, should
lead to interesting nonlinear behavior. Of course, the complicated
situations described here are only obtained for special parameter
values ({\it e.g.} near $v=0$). Generically, we expect the neutral
surface to have only a single minimum.  In particular, this is true
when the group velocity $v$ in (\ref{e:lgl}) is large.

Finally, we examine the possible stability properties of the
spatially-periodic, phase-locked standing-wave solutions of the
Lifshitz equations (\ref{e:lgl}) in the weakly oblique regime. This is
accomplished by deriving the amplitude equations (\ref{e:agl3}) from
(\ref{e:lgl}) for $p\ne 0$.  Inserting the expansion
\bea
A=z_1(t)e^{i(qx+py)}+z_4(t)e^{i(qx-py)}+h.o.t. \nonumber \\
B=z_3(t)e^{-i(qx+py)}+z_2(t)e^{-i(qx-py)}+h.o.t. \label{e:ABexpz}
\eea
into (\ref{e:lgl}), we find that the nonlinear coefficients in
(\ref{e:agl3}) in terms of those in (\ref{e:lgl}) are
\be
a=\tilde{c},\ b=c=f=\tilde{c}+\tilde{g},\ d=2\tilde{c},
\ee
with $\nu$ in (\ref{e:agl3}) replaced by $\nu_{eff}$. It then follows
from our stability calculations in sections III and IV that $PSRe$ are
always {\it unstable} at onset in the weakly oblique regime. Moreover,
$PSRo$ are a stable solution at onset provided $\nu_{eff,r}<0$ and
\be
Re\left\{\nu_{eff}^*(2\tilde {c}+\tilde {g})\right\}> 0. \label{e:PSRolglstab}
\ee
This is precisely the condition that the $PSRo$ bifurcate
supercritically from the trivial solution ({\it cf}. equation
(\ref{e:pswbr})). That $PSRe$ bifurcate unstably close to the Lifshitz
point is consistent with the observation that rectangles are an
unstable solution of the reduced Lifshitz equation (\ref{e:rlgl})
\cite{pk86}, and that $SRe$ are an unstable solution of the full
Lifshitz equations (\ref{e:lgl}) in the absence of forcing ($\mu=0$)
\cite{srk92}.

\section{Comparison with Experiments in Electroconvection}

Before comparing our results with the experiments in
electro-convection of nematics, it is useful to discuss some of the
properties of the unforced experimental liquid crystal system that are
reported in \cite{tr90}. Without temporal forcing the first
instability of the basic state leads to a {\it steady} roll pattern.
Detailed investigation shows, however, strong fluctuations below
threshold which are possibly due to thermal noise.  Crucial in the
present context is the fact that the fluctuations have the character
of (oblique) traveling waves. A natural bifurcation diagram which is
consistent with these findings is shown in fig.\ref{f:ehcbif}; similar
to convection in binary mixtures, there could be a subcritical Hopf
bifurcation which leads to unstable waves which in turn merge with a
branch of steady rolls. In this bifurcation scenario there is a
discontinuous transition from the motionless state directly to finite
amplitude steady rolls when the control parameter is increased beyond
the Hopf bifurcation.

Motivated by experiments on binary-mixture convection, where the
bifurcation to traveling waves is subcritical \cite{rrfts88}, the
effect of resonant periodic forcing on a {\it weakly subcritical} Hopf
bifurcation to one-dimensional traveling waves has been investigated
theoretically \cite{rck91}.  In this subcritical case, it was found
that forcing can lead to stable $PSW$ that arise in a supercritical
bifurcation already below any saddle node bifurcation on the unforced
(as well as of the forced) branch of traveling waves. In particular,
bifurcation diagrams of the kind sketched in fig.\ref{f:pswbifsub}, in
which the parity-breaking bifurcation of the $PSW$ to $TW$ is
subcritical, are possible.  If the unforced traveling waves are
unstable to a large-amplitude steady structure it is quite natural to
assume that this instability persists in the case of sufficiently weak
forcing.  Thus, at the parity-breaking bifurcation to traveling waves
the system could jump from the $PSW$ to the steady structure as
indicated in fig.\ref{f:pswbifsub}.

Without temporal forcing the experimental system has two control
parameters: the amplitude $V$ and frequency $\Omega$ of the applied
ac-voltage. The bifurcation parameter $\nu_r$ depends linearly on the
amplitude $V$, {\it i.e.} $\nu_r\propto (V-V_c)/V_c$. The frequency
$\Omega$ is much greater than the Hopf frequency in the EHC
experiments; our amplitude equations apply to the averaged system. The
nonlinear coefficients in the amplitude equations (\ref{e:agl3})
depend on $\Omega$ and thus can be varied somewhat.  The resonant
temporal forcing introduces two additional control parameters: the
forcing (modulation) amplitude $\Delta V$, or equivalently
$|\mu|^n\propto
\Delta V ^2$ and the detuning $\nu_i =
\omega_e -n \omega_h$ ($n=1,2$).

In the experiments, four regimes have been identified in the
$(\nu_r,\mu)$-plane, as indicated in fig.\ref{f:exp}.  For small
applied voltage $V$ (above the open circles) incoherent traveling-wave
patches are observed. They are interpreted as driven by noise, which
is possibly of thermal origin
\cite{rrtshab91}. This is substantiated by recent measurements of the
spatial Fourier spectra \cite{w93}. Thus the open circles do not
represent a true transition but indicate instead the location where
the instrumental sensitivity is sufficient for the detection of very
weak fluctuations. When the forcing amplitude is increased a
transition to phase-locked standing waves is found (solid triangles).
Depending on the driving frequency $\Omega$ they are either standing
oblique rolls or standing rectangles and can naturally be identified
with the $PSRo$ and $PSRe$, respectively. When increasing the applied
voltage ({\it i.e.} $\nu_r$) further, the $PSRe$ lose stability to a
structure (at the open diamonds) which is also phase locked to the
forcing but is more complicated. During the phase in which the voltage
rises the structure is predominantly a `zig'.  Then the
`zag'-component increases in strength leading to a structure which
resembles rectangles during the peak of the voltage, whereas the
`zag'-component dominates during the decay phase. It is reported that
this structure - in contrast to the $PSRo$ and $PSRe$ - shows no
strong spatial coherence.  We interpret this structure as the $PSCR$
bifurcating off the $PSRe$. Recall that the $PSCR$ is given by
$z_1=z_3=R e^{i\Phi}$, $z_2=z_4=r e^{i\phi}$, so that  a typical
quantity like the vertical velocity $V_z$ is given by ({\it cf}.
(\ref{e:vz}))
\bea
V_z = & 4R\cos(\omega_e t/n + \Phi)\cos(qx+py) + 4r \cos(\omega_e
t/n+\phi)\cos(qx-py) +
h.o.t. \label{e:vzpscr}
\eea
Thus, provided $\Phi -\phi$ is not a multiple of $\pi$, the two
standing waves differ in their phase with respect to the forcing; the
structure alternates between `zigs' (for values of $t$ such that
$\cos(\omega_e t/n+\phi)=0$), `zags' (when $\cos(\omega_e
t/n+\Phi)=0$) and rectangles (when $|R\cos(\omega_e
t/n+\Phi)|=|r\cos(\omega_e t/n+\phi)|$).  Fig. \ref{f:pscr} shows a
representative sequence of $PSCR$ structures. Since the two amplitudes
$r$ and $R$ are different, there exist two symmetry-related $PSCR$
solutions, one in which the `zigs' dominate ($R>r$) and one in which
the `zags' dominate ($r>R$). When the $PSRe$ solution loses stability
to $PSCR$, both zig- and zag-dominated $PSCR$ are expected to arise
and compete with each other, with domains separated by grain
boundaries forming.  It is tempting to attribute the spatial
incoherence observed in the experiments to the existence of such
domains.  Unfortunately, the reported experiments do not provide
enough data to investigate this interesting question.

When the applied voltage is increased further a transition to a steady
structure occurs (to the right of the solid squares) which is
modulated in its amplitude.  This transition is beyond the scope of
the present analysis.  One may speculate that it is related to the
fact that the initial Hopf bifurcation seems to be weakly subcritical
({\it cf}.  fig.\ref{f:ehcbif}): without temporal forcing the system
jumps directly to a steady branch. In the presence of forcing this
could render the parity-breaking bifurcation of the $PSRe$
subcritical. The system would then jump from the $PSCR$ directly to
the steady branch as indicated in fig.\ref{f:ehcbif}.  We have not
studied the stability of the $PSCR$. Since the instability of the
$PSRe$ to $TRe$ leads out of the standing-wave subspace, which
contains also the $PSCR$, one might expect that they inherit that
instability, and consequently, also exhibit a jump-transition to the
steady structure.

The present analysis suggests that the parameter $\rho$, which governs
the competition between the different standing structures, can be made
small by choosing a suitable detuning $\nu_i$ ({\it cf}.
(\ref{e:defrho})).  If this could be achieved below the unforced Hopf
bifurcation ($\nu_r<0$), both transitions from $PSCR$ to $PSRo$ and to
$PSRe$ could be observed at small amplitudes below the onset of the
Hopf bifurcation in the unforced case. In the experimental system the
wave number of the structure is, however, not fixed.  Instead, the
system can choose a wave number from a continuum of values within the
stability limits. In particular, if ${\bf v}=v_x \hat x+ v_y \hat y$
is the group velocity in the oblique regime, then a change $\Delta
q\hat x+\Delta p\hat y$ in the wave vector shifts the detuning to an
effective value: $\nu_{eff,i}=\nu_i + v_x
\Delta q +v_y \Delta p + {\cal O}(\Delta q^2, \Delta p^2, \Delta
q\Delta p)$.  Similarly, near the Lifshitz point, the effective
detuning at the critical wave number $(q_c,p_c=0)$, with $q_c$ given
by (\ref{e:qclif}), is strongly reduced and can remain small over a
range of frequencies; in particular, $\nu_{eff,i}=\nu_i+v
q_c-d_iq_c^2={\cal O}(\epsilon^4)$ \cite{r90a}. This may explain why
in the experiments the location of the transition to $PSCR$, which
depends on the detuning $\nu_i$ or more precisely $\nu_{eff,i}$, did
not change substantially even though the forcing frequency was changed
by almost 50\%.

Finally, our analysis suggests that a transition to $PSCR$ may also be
found if the frequency $\Omega$ is chosen such that the first
transition is not to rectangles but to standing oblique rolls
($PSRo$), as long as $\rho$ is positive and not too large in the
regime where the $PSRo$ arise. Experimentally this has not been tested
in detail.

\section{Conclusion}

In summary, we have studied the effect of a resonant temporal forcing
on oblique traveling waves close to onset in an anisotropic system. We
have found that such a forcing can excite various standing wave
structures which are phase locked to the forcing: rolls, rectangles
and cross rolls. The latter arise in secondary bifurcations from
either the phase-locked roll or rectangle structures, and in many
cases (if not all) provide a transition between them. These
bifurcations depend strongly on external control parameters such as
the forcing frequency (effective detuning). In particular, if the wave
number of the structure could be fixed, one would be able to choose
the control parameters such that a wide variety of bifurcation
scenarios could be observed. In the general case, however, the
attainable effective detuning is limited. Nevertheless, by adjusting
the frequency $\Omega$ of the applied ac-voltage in the unforced case,
so that the system is close to a transition between standing rolls and
rectangles, the available detuning may be sufficient to observe the
transition from phase-locked standing rolls to phase-locked standing
rectangles via the general cross-rolls solution.

A striking feature of the standing rolls and rectangles observed in
the experiments is their strong spatial coherence. This is perhaps not
surprising since they arise through bifurcations in the amplitude
equations that involve only real eigenvalues. Close to onset, the
standing rolls can be modeled by a single Ginzburg-Landau equation
with real coefficients. This equation admits a stable band of wave
numbers and exhibits diffusive phase dynamics; moreover, it can be
derived from a potential.  Similarly, we expect the standing
rectangles to be described by two coupled Ginzburg-Landau equations
with real coefficients. We propose that the spatial incoherence
associated with the standing cross rolls may be due to the formation
of grain boundaries separating domains of symmetry-related structures.
A more detailed experimental, as well as theoretical, investigation of
this regime would be interesting. Such studies could reveal whether
this interpretation is correct and, if so, identify the dynamics
associated with the domain walls.

In the vicinity of a Lifshitz point, we model the system by two
coupled complex generalized Ginzburg-Landau equations which take into
account the spatial degrees of freedom. We find, in this regime, that
the phase-locked standing rectangles are always unstable.
Phase-locked standing rolls are the preferred spatially-periodic
solution in this regime.

Finally, the results presented in this paper may also be pertinent to
parametrically forced systems which do not exhibit a Hopf bifurcation,
but which do support weakly damped waves. This, for instance, is the
case in the Faraday experiment for low-viscosity fluids.

\acknowledgments
We would like to thank M.~de la Torre Ju\'arez and I.~Rehberg for
discussing their experiments with us, and E. Knobloch for useful
discussions on subjects related to this paper.  This work was
supported by a NATO collaborative research grant \#900276. H.R.
acknowledges support from the NSF/AFOSR under grant number DMS-9020289
and support by DoE grant DE-FG02-92ER14303.  M.S. and H.R. would like
to thank the University of Bayreuth, where much of this work was done,
for its hospitality.  Part of this work was also carried out at the
Institute for Theoretical Physics at the University of California,
Santa Barbara, which is supported by NSF grant PHY89-04035.

\vspace{.3cm}
\noindent
$^+$ Also at the Institute for Theoretical Physics, University of California,
Santa Barbara, CA 93106, USA.\\
$^\#$ present address: Department of Engineering Sciences and Applied
Mathematics,
Nortwestern University, Evanston, IL 60208

\begin{figure}
\caption{Phase diagram for a periodically forced supercritical
Hopf bifurcation to traveling waves. Phase-locked standing waves
bifurcate from the motionless state on the line $PSW$, traveling waves
bifurcate at the line marked $H$. The standing waves undergo a
parity-breaking bifurcation at $PB$, a saddle-node bifurcation at $SN$
and a Hopf bifurcation to unlocked standing waves at $SW$.
\protect{\label{f:pdsup}}
}
\end{figure}

\begin{figure}
\caption{Typical bifurcation diagram for a periodically forced supercritical
Hopf bifurcation to traveling waves as obtained along the dotted line
in fig.\protect{\ref{f:pdsup}}.
\protect{\label{f:pswbif}}
}
\end{figure}

\begin{figure}
\caption{Sketch of the bifurcation
diagram of $PSRo$, $PSRe$ and $PSCR$ for $\Delta \mu >0$ and $\rho >0$
in case I.
\protect{\label{f:psrpsba}}
}
\end{figure}

\begin{figure}
\caption{Sketch of the bifurcation diagram
of $PSRo$, $PSRe$ and $PSCR$ for $\Delta \mu <0$ and $\rho >0$ in case
I.
\protect{\label{f:psrpsbb}}
}
\end{figure}

\begin{figure}
\caption{Neutral surface for the forced Lifshitz equation
(\protect{\ref{e:lgl}}) with
$\nu_r=-0.2$, $d=1+i$, $s=-1+0.585i$, $h=-4.27+2.5i$ and
$\nu_i=-0.34$. In the nonlinear regime zig-zag structures are expected
to arise. The local wave number of the zigs and of the zags lies in
one of the three minima of the neutral surface.
\protect{\label{f:ns1}}
}
\end{figure}

\begin{figure}
\caption{Neutral surface for $\nu_i=-0.3$. Other parameters as in
fig.\protect{\ref{f:ns1}}.
\protect{\label{f:ns2}}
}
\end{figure}

\begin{figure}
\caption{Neutral surface for $d_i=-0.5$. Other
parameters as in fig.\protect{\ref{f:ns2}}.
\protect{\label{f:ns3}}
}
\end{figure}

\begin{figure}
\caption{Neutral surface for $d_i=-0.7$. Other parameters as in
fig.\protect{\ref{f:ns2}}.
\protect{\label{f:ns4}}
}
\end{figure}

\begin{figure}
\caption{Sketch of the
bifurcation diagram suggested by experiments on electro-convection in
nematic liquid crystals. \protect{\label{f:ehcbif}} }
\end{figure}

\begin{figure}
\caption{Effect of a periodic forcing on the
subcritical Hopf bifurcation to traveling waves sketched in
fig.\protect{\ref{f:ehcbif}}. The parity-breaking bifurcation of the $PSW$ to
$TW$ can be subcritical, thus inducing a jump-transition to the steady
structure.
\protect{\label{f:pswbifsub}}
}
\end{figure}

\begin{figure}
\caption{Experimental
phase diagram for temporally forced convection in nematic liquid
crystals, reproduced from \protect{\cite{tr90}}. (See the text for a discussion
of the various regions.) a) $\Omega = 0.50 Hz$, b) $\Omega =0.31 Hz$.
\protect{\label{f:exp}}
}
\end{figure}

\begin{figure}
\caption{Typical time sequence for a representative
phase-locked standing cross rolls solution ($PSCR$). Note that the
structure alternates between a zig- and a zag-dominated state.
\protect{\label{f:pscr}}
}
\end{figure}

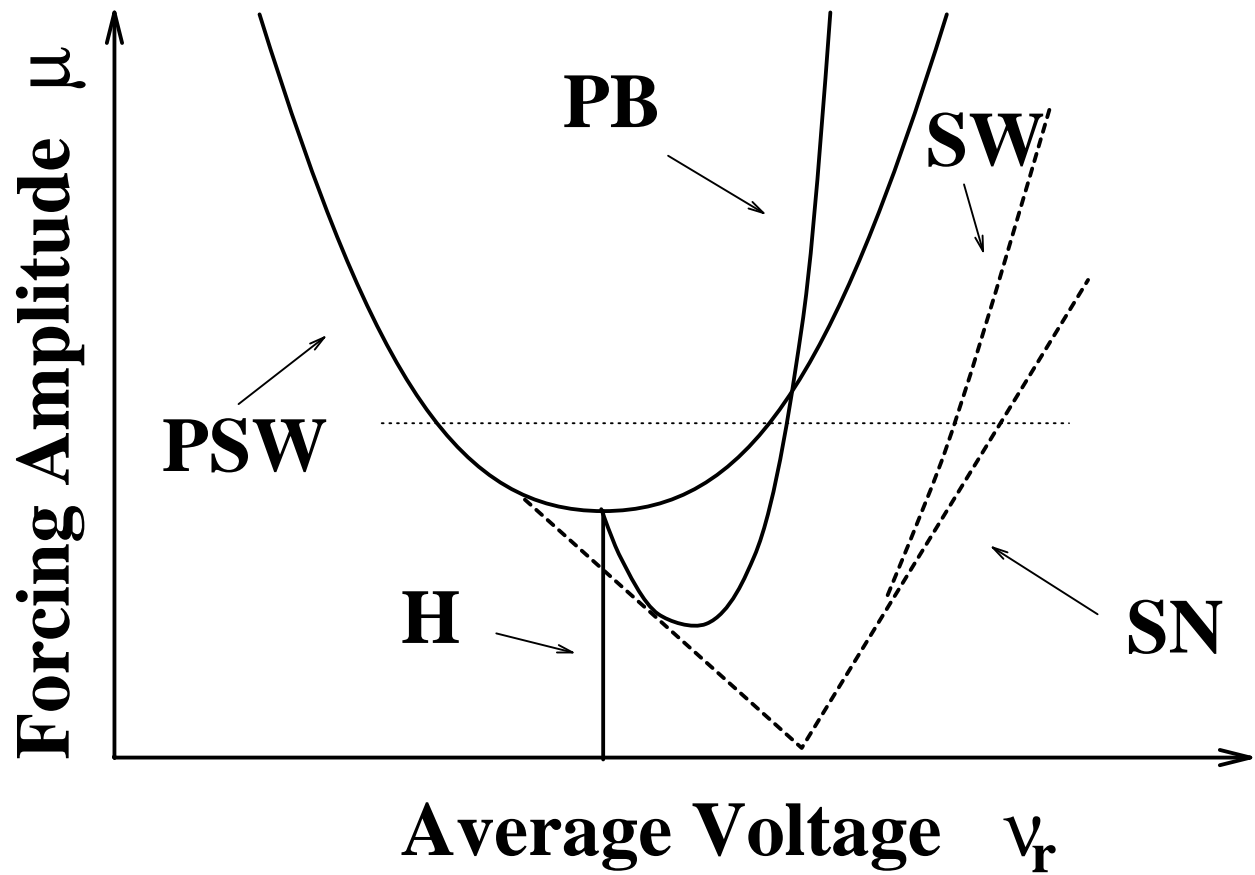
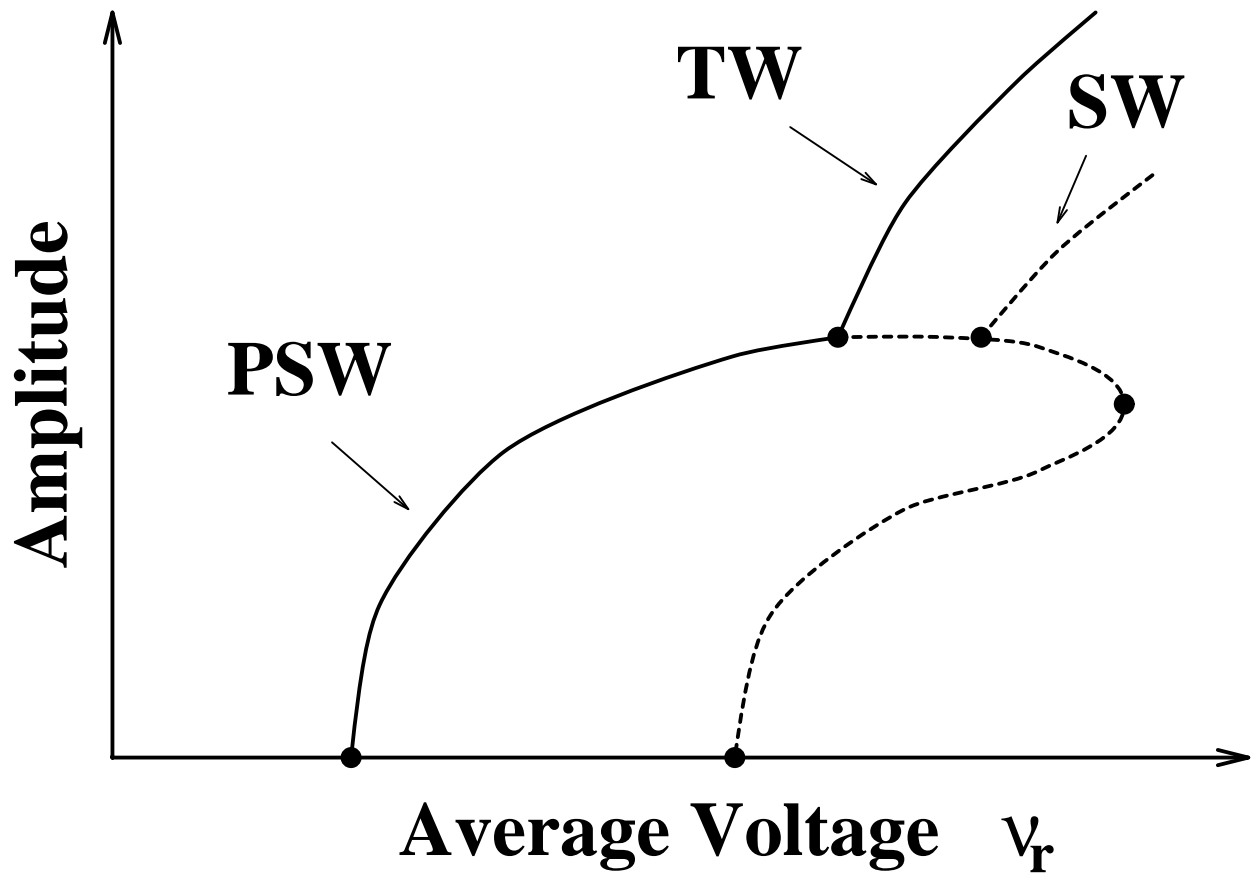
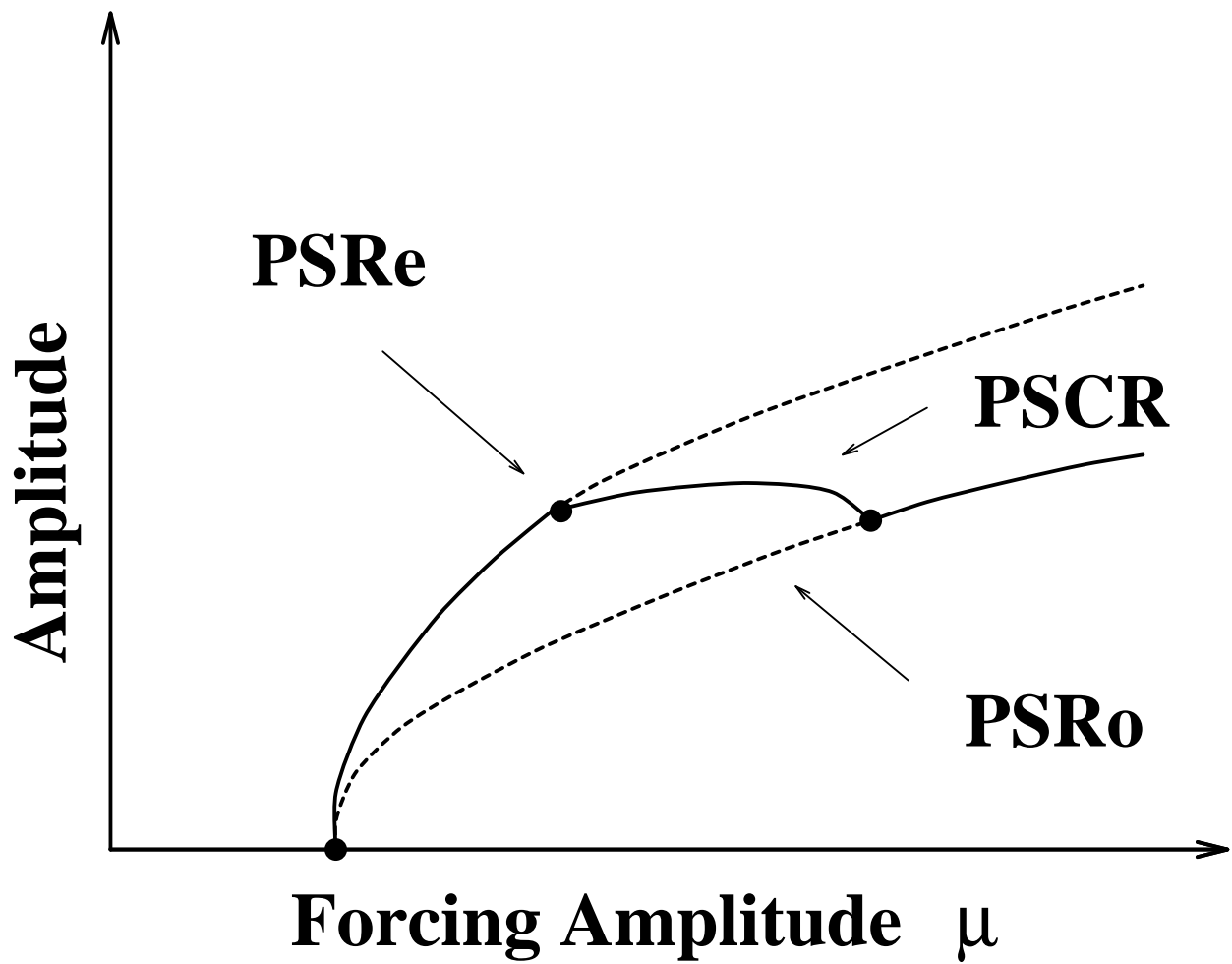
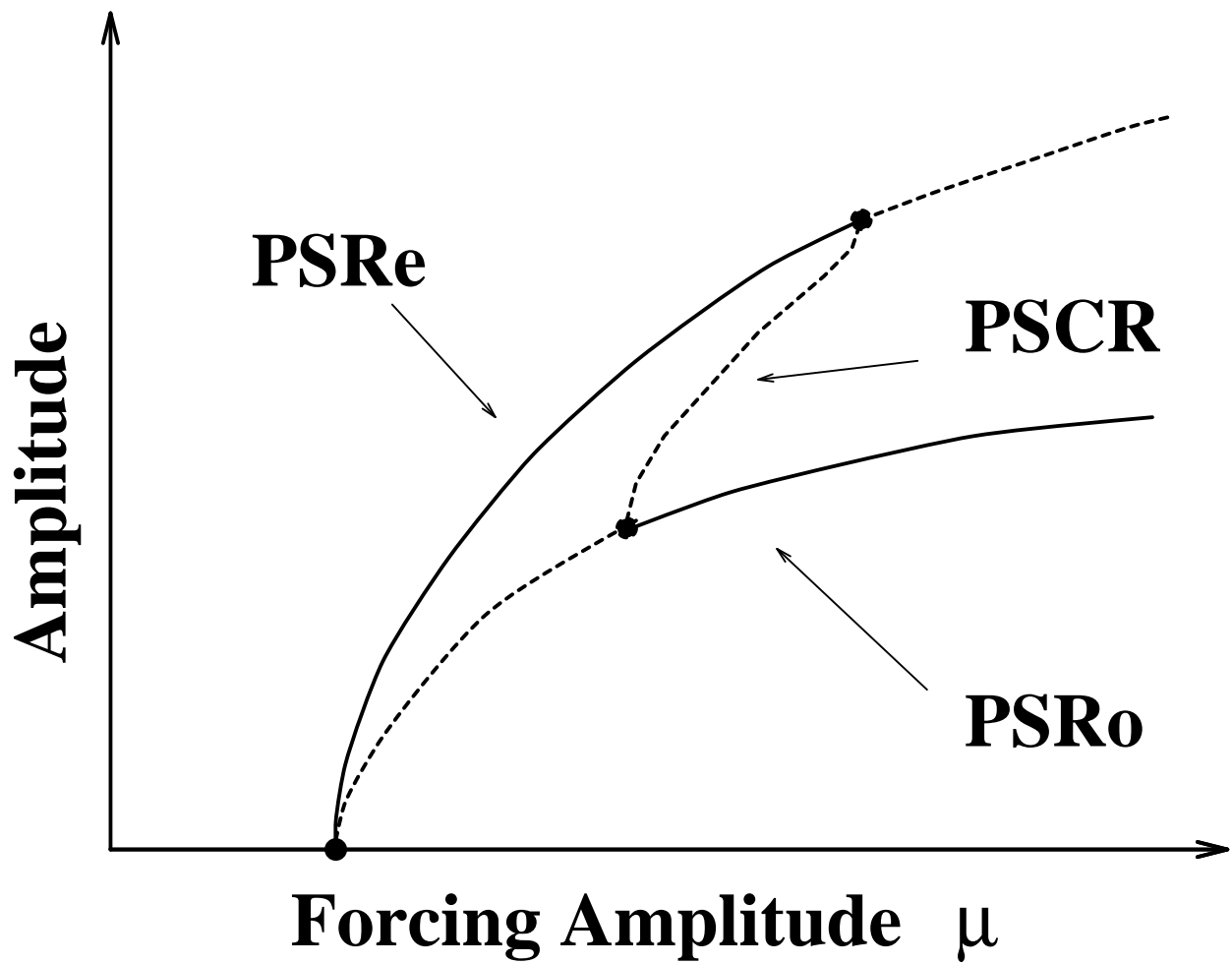
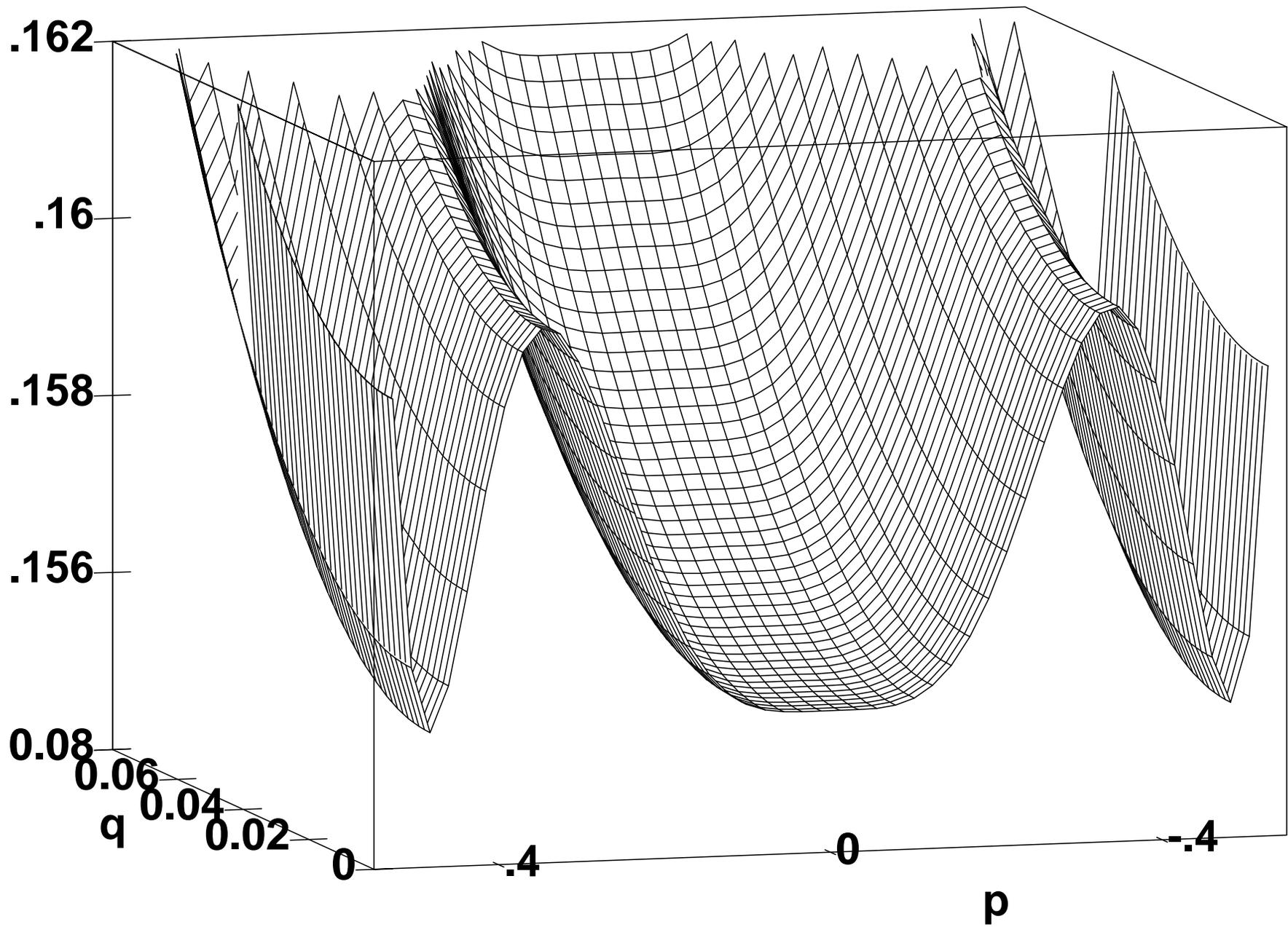
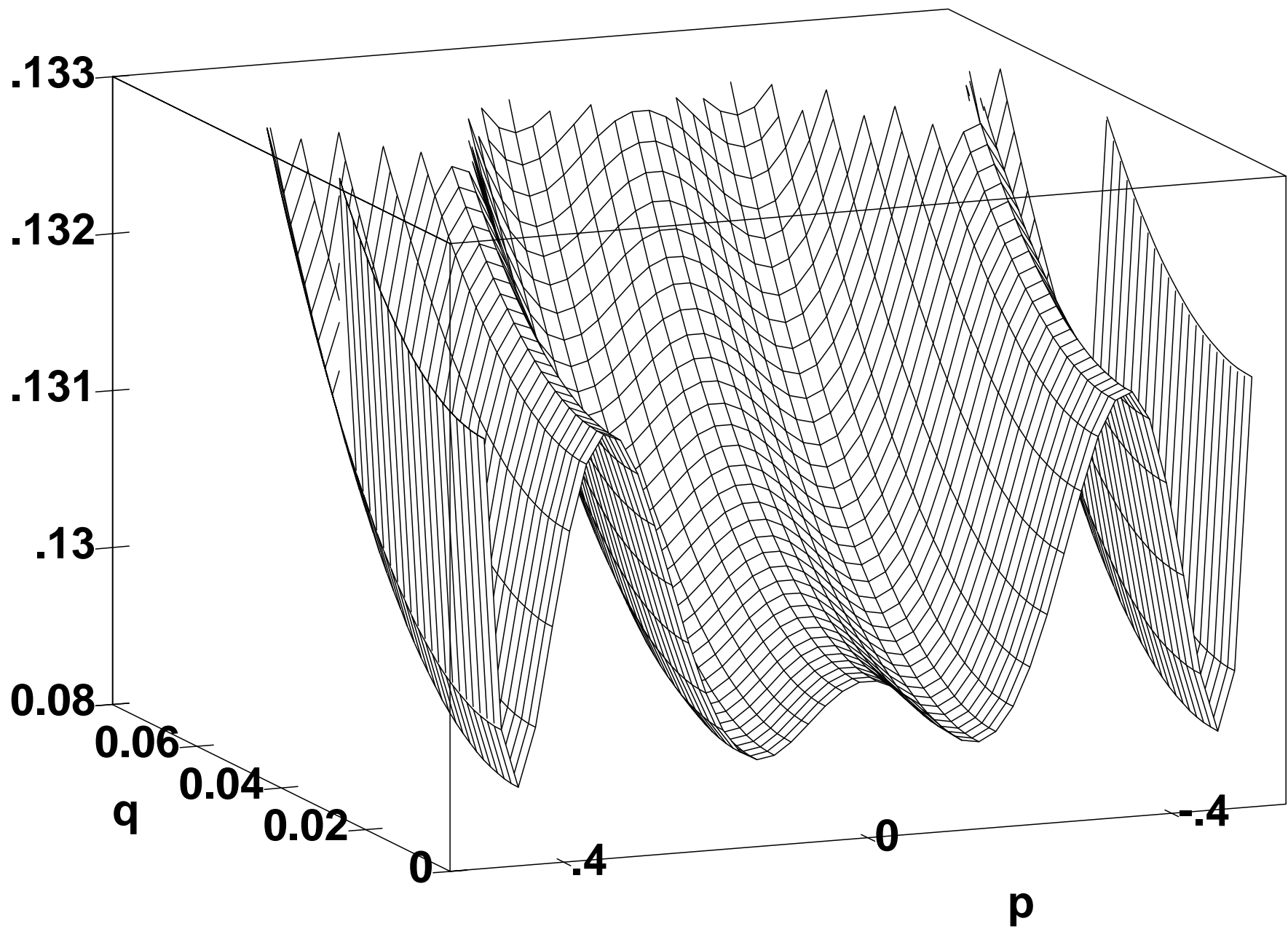
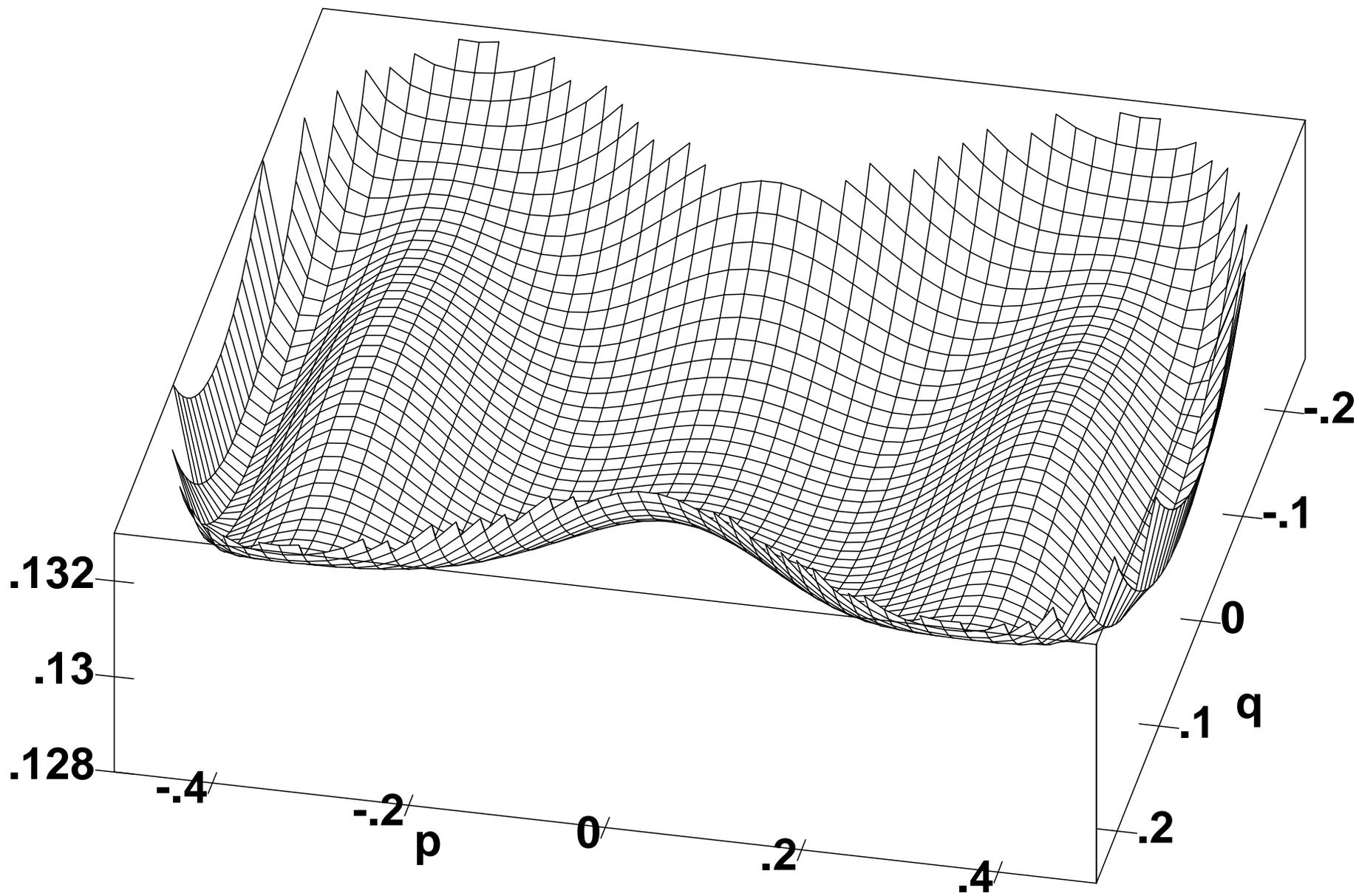
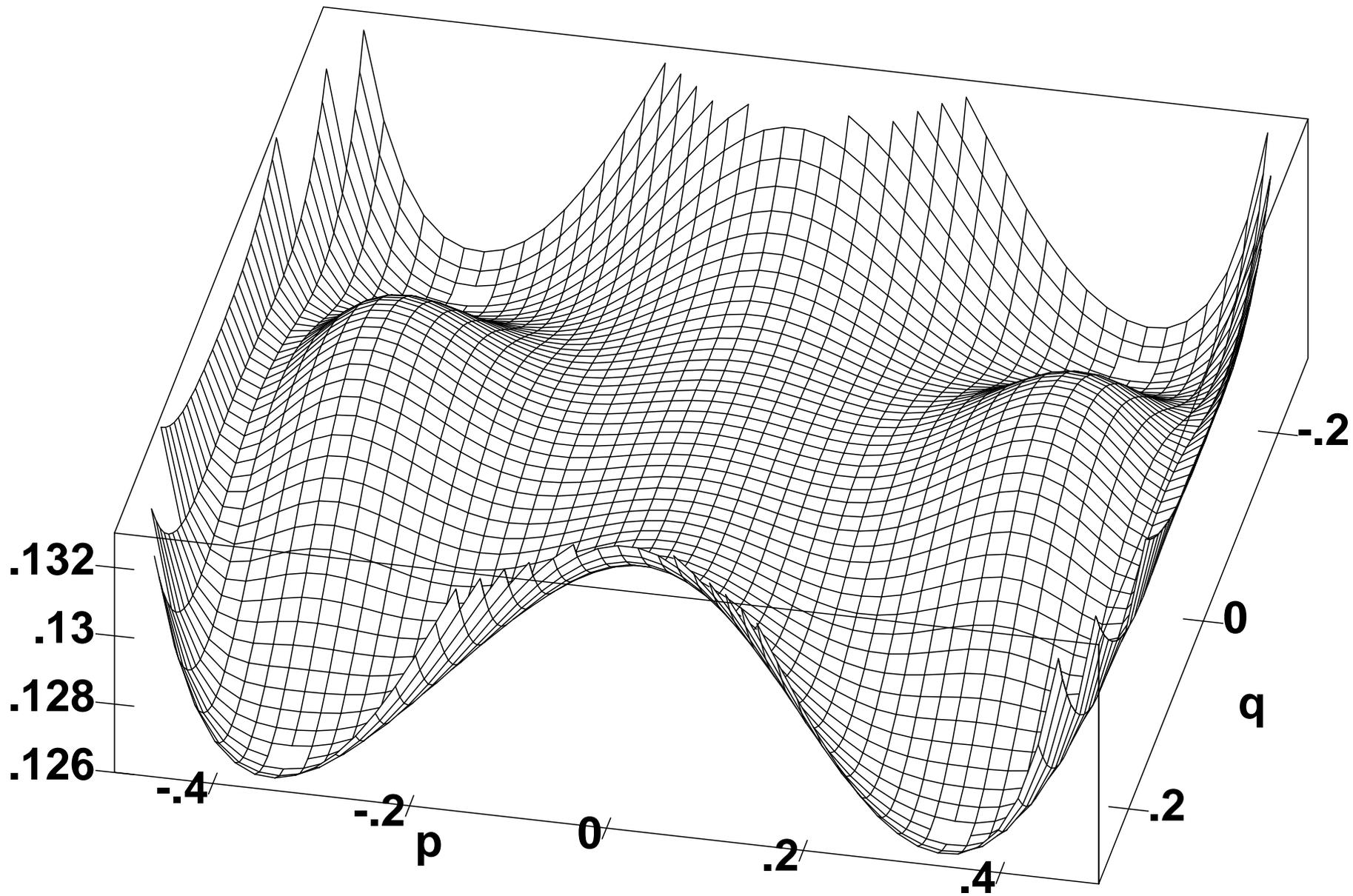
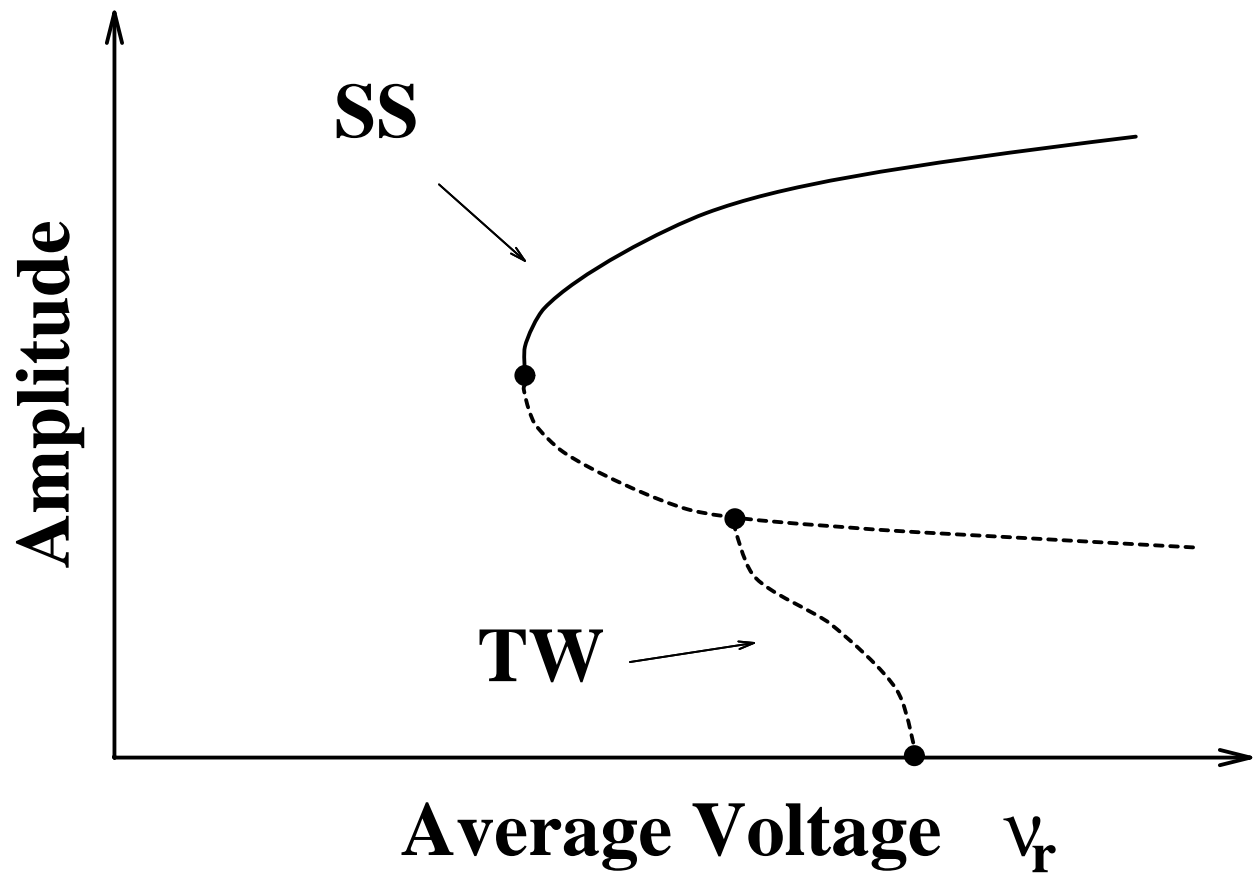
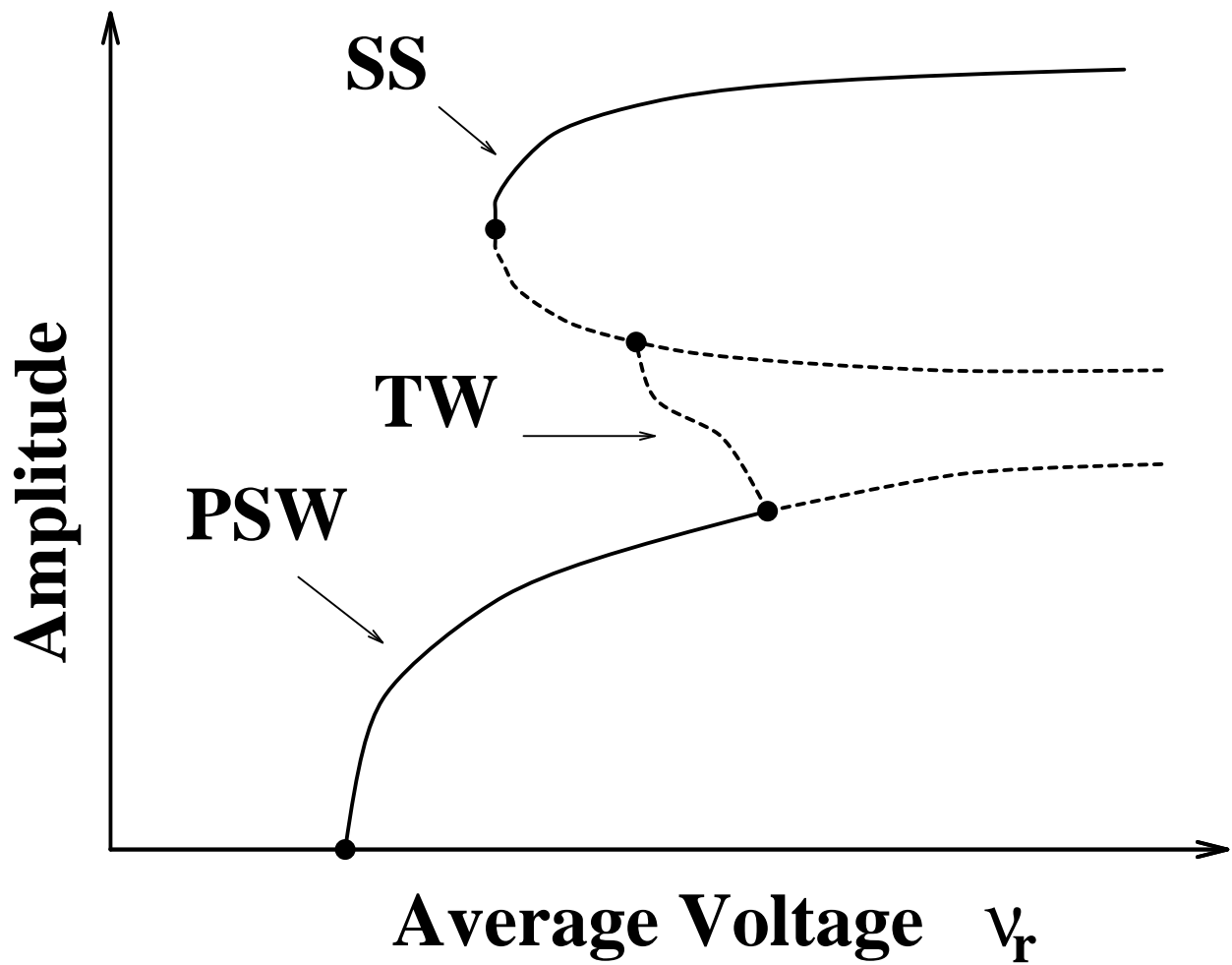
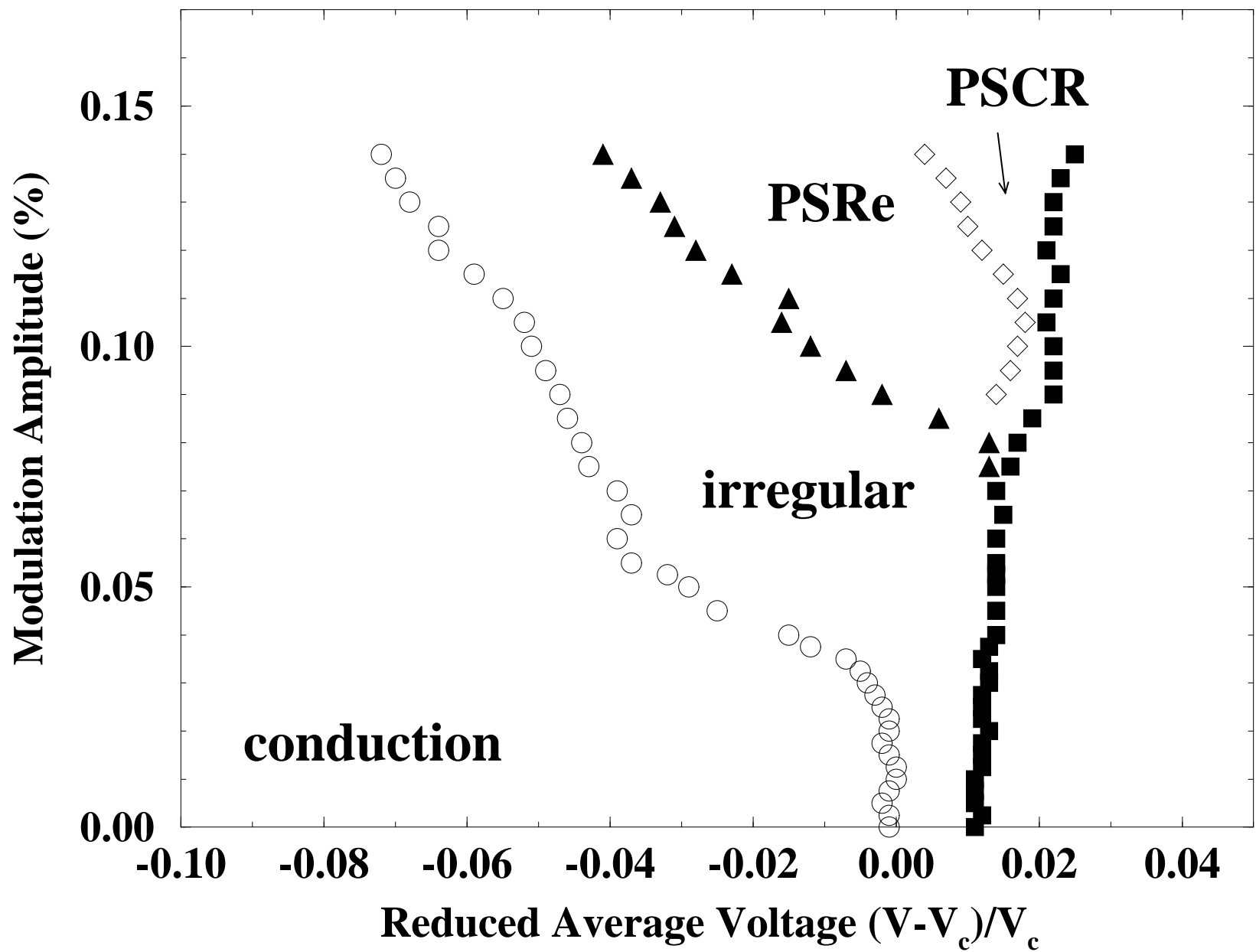
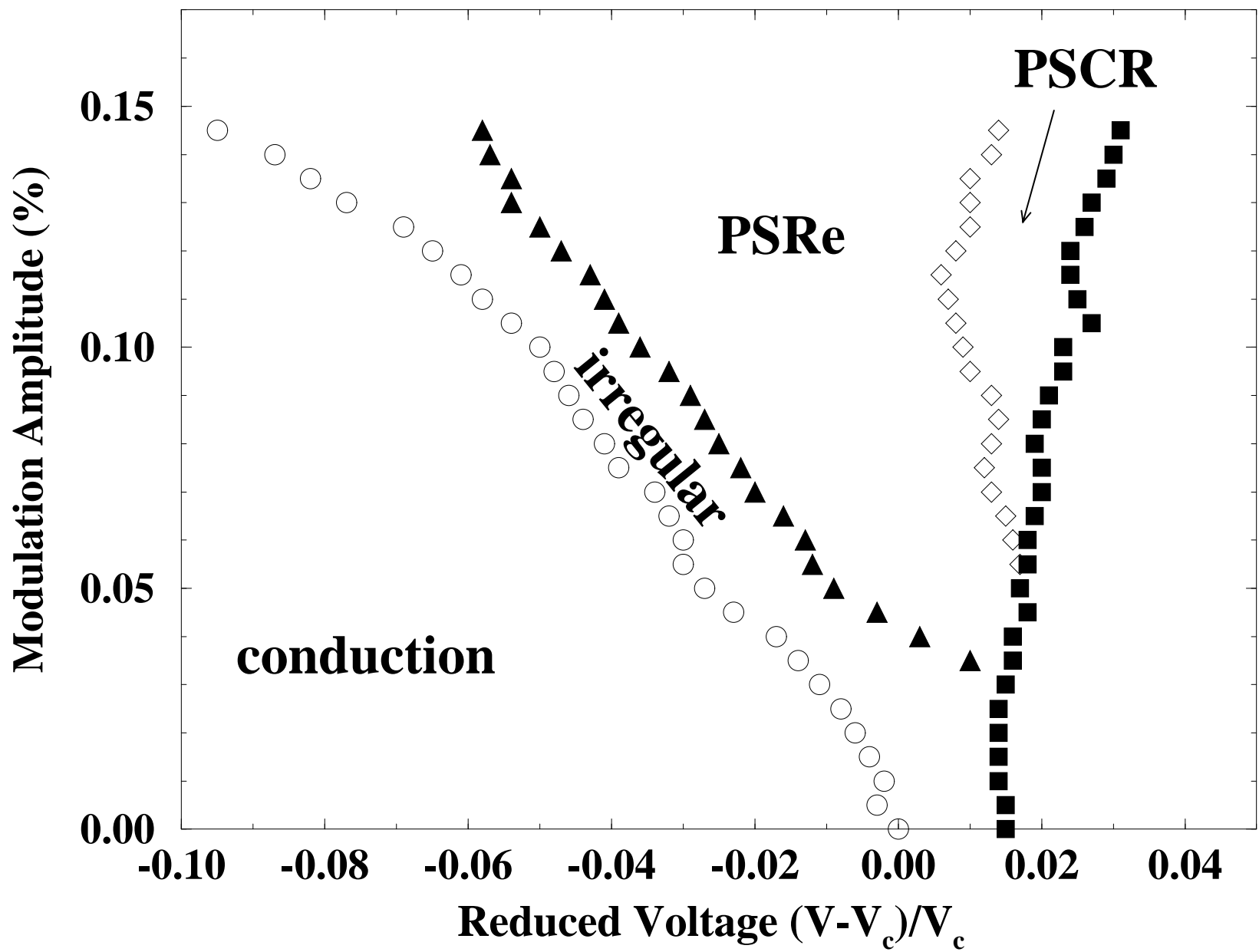
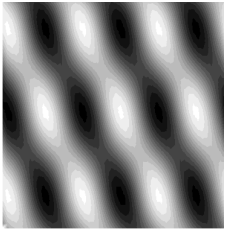
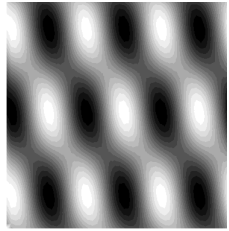
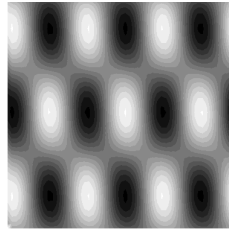
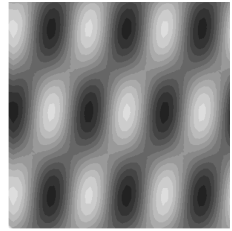
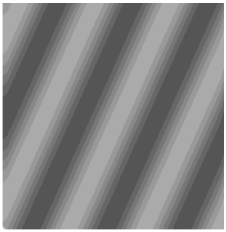
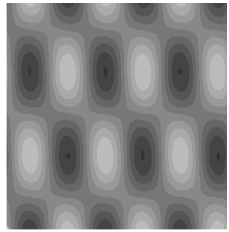
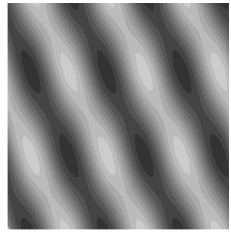
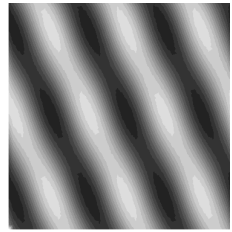
\end{document}